\documentclass{jfm}
\usepackage{graphicx}
\usepackage{epstopdf, epsfig}
\usepackage{color}

\shorttitle{Instabilities of a circular Couette flow with radial temperature gradient}
\shortauthor{O. N. Kirillov, I. Mutabazi}

\title{Short wavelength local instabilities of a circular Couette flow with radial temperature gradient}

\author{Oleg N. Kirillov\aff{1}
  \corresp{\email{kirillov@mi.ras.ru}}
 \and Innocent Mutabazi\aff{2}}

\affiliation{\aff{1}Steklov Mathematical Institute, Russian Academy of Sciences,
Gubkina 8, Moscow 119991, Russia
\aff{2}Laboratoire Ondes et Milieux Complexes (LOMC), UMR 6294, CNRS-Universit\'e du Havre, Normandie Universit\'{e}, B.P. 540, 76058 Le Havre Cedex, France}

\begin{document}
\newcommand{\eps}{{\varepsilon}}
\newcommand{\proofend}{$\Box$\bigskip}
\newcommand{\C}{{\mathbf C}}
\newcommand{\Q}{{\mathbf Q}}
\newcommand{\R}{{\mathbf R}}
\newcommand{\Z}{{\mathbf Z}}
\newcommand{\RP}{{\mathbf {RP}}}
\newcommand{\CP}{{\mathbf {CP}}}
\newcommand{\A}{{\rm Area}}
\newcommand{\Le}{{\rm Length}}

\newcommand{\be}[1]{\begin{equation}\label{#1}}
\newcommand{\ee}{\end{equation}}
\newcommand{\ba}[1]{\begin{eqnarray}\label{#1}}
\newcommand{\ea}{\end{eqnarray}}
\newcommand{\rf}[1]{(\ref{#1})}
\newcommand{\nn}{\nonumber}
\newcommand{\red}{\textcolor{red}}
\newcommand{\blue}{\textcolor{blue}}
\newcommand{\green}{\textcolor{green}}

\maketitle

\begin{abstract}
We perform a linearized local stability analysis for short-wavelength perturbations of a circular Couette flow with the radial temperature gradient. Axisymmetric and nonaxisymmetric perturbations are considered and both the thermal diffusivity and the kinematic viscosity of the fluid are taken into account. The effect of the asymmetry of the heating both on the centrifugally unstable flows and on the onset of the instabilities of the centrifugally stable flows, including the flow with the Keplerian shear profile, is thoroughly investigated. It is found that the inward temperature gradient destabilizes the Rayleigh stable flow either via Hopf bifurcation if the liquid is a very good heat conductor or via steady state bifurcation if viscosity prevails over the thermal conductance.
\end{abstract}


\section{Introduction}

Circular Couette flow of a viscous Newtonian fluid between two coaxial differentially rotating cylinders is a canonical system for modeling instabilities leading to spatio-temporal patterns and transition to turbulence in many natural and industrial processes. Modern astrophysical applications require understanding of basic instability mechanisms in rotating flows with the Keplerian shear profile in accretion- and protoplanetary disks that are hydrodynamically stable according to the centrifugal Rayleigh criterion. Usually, these instabilities are a consequence of additional factors such as electrical conductivity of the fluid and magnetic field of the central gravitating object \citep{Chandrasekhar,L1987,FV1995,UB1998,KK2013,KSF2014,CKH2015,BP2016}. However, in the so-called `dead-zones'
of protoplanetary disks that are characterized by high electrical resistivity due to very low ionization levels, some magnetohydrodynamic (MHD) instabilities become inefficient (e.g. the standard Velikhov-Chandrasekhar magnetorotational instability (MRI) in an axial magnetic field). Alternative mechanisms include MRIs caused by azimuthal or helical magnetic fields that work well also in the inductionless limit at a very low conductivity  \citep{KS2013,KSF2014} as well as pure hydrodynamic finite-amplitude nonlinear instabilities \citep{Zombie2013}. Some recent studies indicated a possibility that Keplerian disks with strong mean radial temperature gradients can support the so-called Goldreich-Schubert-Fricke (GSF) instability, which is the instability of short radial wavelength inertial modes \citep{EM1980,UB1998,NGU2013,BP2016}. These astrophysical implications renewed interest to the effect of radial temperature gradient on the stability of a circular Couette flow and its observation in the laboratory \citep{YNM2013,MYM2015}.

Typically, instabilities in the circular Couette flow are studied numerically, for instance, via normal mode analysis or direct numerical simulations. An alternative technique developed in the works \citep{E1981,L1987,L1991,LH1991,DS92,LH1993,EY1995,FV1995}
is based on the perturbations of a background flow in a small parameter, representing short wavelength, as in geometric optics \citep{LSB1996}. These perturbations are localized
wave envelopes moving along the trajectories of fluid elements. The flow is unstable if the amplitude of the envelope demonstrates an unbounded growth in time along at least one trajectory.

The geometric optics stability analysis requires the representation of the solution to the governing equations linearized about the background flow as a generalized progressive wave expansion \citep{E1981}. Since
the leading order term in this expansion dominates the solution for a sufficiently long time, provided that the short wavelength parameter is small enough, it is sufficient for detecting instability to determine the growth rate of the lowest-order amplitude  \citep{E1981,L1991}. The growth rates are given by real parts of roots of the dispersion relation of the linear transport equation for the lowest-order amplitude that depends on the wave vector of the perturbation. The wave vector evolves along a stream line of the flow and this evolution is governed by an eikonal equation. The stream line passing through a particular point fulfills a corresponding trajectory equation. Thus, detecting instabilities localized near a particular fluid element location, moving with the flow, reduces to solving a system of ordinary differential equations for the wave vector and amplitude, along particle paths in the underlying flow, with coefficients depending on the unperturbed velocity field \citep{LSB1996}.

The geometric optics stability analysis has proved successful in solving the problems of ideal hydrodynamics and magnetohydrodynamics related to stability of rotating flows of an incompressible fluid \citep{E1981,L1987,L1991,LH1991,LH1993,FV1995,LSB1996,KK2013}. It was extended to the viscous and resistive case in \citep{L1991,DS92,EY1995,K2013,KSF2014,AS2015}.
In particular, this technique allowed one to find analytically the neutral stability surfaces in the parameter space, the frequencies of the unstable modes, and their growth rates that agreed well with the numerical simulations \citep{EY1995,LSB1996,KSF2014,CKH2015,SK2015}.

In this paper we apply the geometric optics stability analysis to the circular Couette flow of a viscous Newtonian fluid with the temperature gradient in the absence of gravity, but retaining the term of the centrifugal buoyancy. We derive a system of characteristic equations that includes the transport equations for the lowest-order amplitude of the envelope of the localized perturbation and find the dispersion relation that takes into account the radial variation of the angular velocity and the temperature as well as the kinematic viscosity and the thermal diffusivity.

Using algebraic stability criteria for localization of roots of polynomials in the left half of the complex plane, we obtain two stability conditions in compact and explicit form; one of them generalizes the Rayleigh discriminant for stationary axisymmetric instabilities to include the viscosity effects, and the other provides marginal stability curve in the parameter plane for oscillatory instabilities. Codimension-2 points on the marginal stability thresholds are identified and the growth rates and frequencies of the instabilities are found analytically. The theoretical results are applied to the cylindrical Couette flow with the parameters evaluated at the geometric mean radius both in the Rayleigh unstable and in the Rayleigh stable regimes. In the case of the sole inner cylinder rotating we confirm and extend the numerical results of \cite{MYM2015} and provide analytical expressions for the onset of oscillatory and stationary instabilities, the coordinates of the codimension-2 points, and the Hopf frequency. In the case of the sole outer cylinder rotation, solid body rotation, and rotating flow with Keplerian shear we find the destabilizing effect of the inward temperature gradient that leads to the oscillatory instability at small values of the Prandtl number ($\rm Pr<1$) and to the stationary instability at ${\rm Pr}>1$.

\section{Equations of motion}

We consider the flow of an incompressible viscous fluid of the density $\rho$, the kinematic viscosity $\nu$  and the thermal diffusivity  $\kappa$ in the presence of a radial temperature gradient applied to a differentially rotating  cylindrical annulus.  The ratio ${\rm Pr}=\nu/\kappa$ constitutes the Prandtl number.

The governing equations written in the inertial frame of reference read \citep{YNM2013,MYM2015}:
\ba{nsht}
\nabla \cdot {\bf u}&=&0, \\
\frac{\partial {\bf u}}{\partial t}+{\bf u} \cdot \nabla {\bf u}+\frac{1}{\rho}\nabla p - \nu \Delta {\bf u}+\alpha \theta {\bf g}_c&=&0,  \\
\frac{\partial \theta}{\partial t}+{\bf u} \cdot \nabla {\theta}-\kappa \Delta \theta &=&0,
\ea
where $p$ is the pressure and  $\theta$ is the temperature deviation from a reference temperature.  The parameter
$\alpha$ is the coefficient of thermal expansion.

The flow equations are written in cylindrical coordinates $(r,\varphi,z)$ in which  the velocity field  ${\bf u}=(u,v,w)^T$, where the superscript $T$ denotes transposition. The term ${\bf u} \cdot \nabla {\bf u}$ contains the centrifugal-like acceleration ${\bf g}_c=(v^2/r,0,0)^T$ and the Coriolis-like acceleration $(0,uv/r,0)^T$ coming from the geometry. The Oberbeck-Boussinesq approximation \citep{Chandrasekhar} is used, i.e. we assume small variations of the density  with the temperature only in the centrifugal force term leading to the centrifugal buoyancy while the other fluid properties ($\nu, \kappa, \alpha$) are kept constant. This approximation allows one to keep the flow incompressibility condition and eliminates the acoustic modes from the problem analysis. The application of the Oberbeck-Boussinesq approximation to rotating flows can be found in \citep{Lopez2013}. The gravity $g$ is neglected in order to focus only on the destabilization effects of the centrifugal buoyancy ${\alpha \theta \bf g}_c$.

We assume that the cylindrical annulus has an infinite length. The system of equations \rf{nsht} possesses a stationary axisymmetric solution describing the base flow state: ${\bf u}_B=(0,V(r) = r \Omega(r),0)^T$, $p_B=P(r)$, and $\theta_B=\Theta(r)$. Consider a perturbation of this base flow state: ${\bf u}={\bf u}_B+\widetilde{\bf u}$, $p=p_B+\widetilde p$, and $\theta=\theta_B+\widetilde \theta$.
Then the equations of the flow perturbations linearized around  the base flow state read
\ba{le}
\nabla \cdot \widetilde {\bf u}&=&0,\\
\frac{\partial \widetilde{\bf u}}{\partial t} +{\bf u}_B \cdot \nabla \widetilde{\bf u} + \widetilde{\bf u} \cdot \nabla {\bf u}_B-\nu\Delta \widetilde{\bf u}+\frac{1}{\rho}\nabla \widetilde p+\alpha \theta_B \widetilde {\bf g}_c+\alpha \widetilde \theta {\bf g}_{cB}&=&0,\\
\frac{\partial \widetilde \theta}{\partial t} +{\bf u}_B \cdot \nabla \widetilde{\theta} +\widetilde{\bf u} \cdot \nabla \theta_B-\kappa \Delta \widetilde \theta&=&0,
\ea
with the basic centrifugal gravity ${\bf g}_{cB}=(r\Omega^2,0,0)^T$ and the perturbative centrifugal gravity $\widetilde {\bf g}_c=(2\Omega \widetilde v,0,0)^T$
so that we can write the centrifugal
buoyancy terms as follows
\ba{force}
\alpha \theta_B \widetilde {\bf g}_c&=&2\Omega \alpha \theta_B {\bf e}_r {\bf e}_{\varphi}^T \widetilde{\bf u},\\
\alpha \widetilde \theta {\bf g}_{cB}&=&\alpha \widetilde \theta r\Omega^2 {\bf e}_r,
\ea
where ${\bf e}_r$ and ${\bf e}_{\varphi}$ are the radial and azimuthal unit vectors, respectively.

The gradients of the background fields are
\be{gbf}
\nabla{\bf u}_B=\Omega\left(
                                  \begin{array}{ccc}
                                    0 & -1 & 0 \\
                                    1+2{\rm Ro} & 0 & 0 \\
                                    0 & 0 & 0 \\
                                  \end{array}
                                \right),
                                \quad
\nabla \theta_B=\left(
                  \begin{array}{c}
                    D\Theta \\
                    0 \\
                    0 \\
                  \end{array}
                \right),
\ee
where \be{ro}{\rm Ro}=\frac{r D\Omega}{2\Omega}\ee is the Rossby number and $D = d/dr$. Denoting $\mathcal{U}=\nabla{\bf u}_B$, we formulate the linearized equations of motion in the final form:
\ba{le2}
\nabla \cdot \widetilde{\bf u}&=&0,\\
\left(\frac{\partial }{\partial t} +\mathcal{U}+2\Omega \alpha \theta_B {\bf e}_r {\bf e}_{\varphi}^T+{\bf u}_B \cdot \nabla  \right)\widetilde{\bf u}-\nu\Delta \widetilde{\bf u}+\frac{1}{\rho}\nabla \widetilde p+\alpha  r\Omega^2 {\bf e}_r \widetilde \theta&=&0,\label{le2b}\\
\left(\frac{\partial }{\partial t} +{\bf u}_B \cdot \nabla\right) \widetilde{\theta} +\widetilde{\bf u} \cdot \nabla \theta_B-\kappa \Delta \widetilde \theta&=&0.\label{le2c}
\ea

\section{Evolution of the localized perturbations}

Let $\epsilon$ be a small parameter ($0<\epsilon \ll 1$). We are looking for a solution of the linearized equations \rf{le2}, \rf{le2b}, and \rf{le2c} in the form of the  generalized progressive wave expansions \citep{E1981}
\ba{go}
\widetilde{\bf u}&=&\left({\bf u}^{(0)}({\bf x},t)+\epsilon {\bf u}^{(1)}({\bf x},t)\right){\exp}\left({{i{\epsilon}^{-1} \Phi({\bf x},t)}}\right)+\epsilon {\bf u}^{(r)}({\bf x},t,\epsilon),\\
\widetilde{\theta}&=&\left({\theta}^{(0)}({\bf x},t)+\epsilon {\theta}^{(1)}({\bf x},t)\right){\exp}\left({{i{\epsilon}^{-1} \Phi({\bf x},t)}}\right)+\epsilon {\theta}^{(r)}({\bf x},t,\epsilon),\label{go2}\\
\widetilde{p}&=&\left({p}^{(0)}({\bf x},t)+\epsilon {p}^{(1)}({\bf x},t)\right){\exp}\left({{i{\epsilon}^{-1} \Phi({\bf x},t)}}\right)+\epsilon {p}^{(r)}({\bf x},t,\epsilon),
\ea
where $\Phi$ is generally a complex-valued scalar function that represents the phase of the wave or the eikonal and  the remainder terms ${\bf u}^{(r)}, \theta^{(r)}, p^{(r)}$ are assumed to be uniformly bounded in $\epsilon$ on any fixed time interval \citep{E1981,L1991,LH1991,LH1993,LSB1996}.

\cite{Maslov1986} observed that high frequency oscillations ${\exp}\left({{i{\epsilon}^{-1} \Phi({\bf x},t)}}\right)$ quickly die out because of viscosity unless one assumes quadratic dependency of viscosity on the small parameter $\epsilon$. Hence, following \citep{Maslov1986,DS92,KSF2014,AS2015}, we assume that $\nu=\epsilon^2 \widetilde \nu$ and $\kappa=\epsilon^2 \widetilde \kappa$.

Substituting the asymptotic series \rf{go} to the incompressibility condition \rf{le2} and collecting terms at $\epsilon^{-1}$ and $\epsilon^{0}$, we find
\ba{sc}
\epsilon^{-1}:\quad\quad\quad\quad\quad~~\, {\bf u}^{(0)}\cdot \nabla \Phi &=&0,\\
\epsilon^{0}:\quad \nabla \cdot {\bf u}^{(0)}+i {\bf u}^{(1)}\cdot\nabla \Phi &=&0.
\ea
Similar procedure applied to \rf{le2b} and \rf{le2c}
yields the two systems of equations
\ba{first}
&\epsilon^{-1}:~~~~~~ \left(
  \begin{array}{cc}
    \frac{\partial \Phi}{\partial t}  +{\bf u}_B \cdot \nabla \Phi & 0 \\
    0 & \frac{\partial \Phi}{\partial t} +{\bf u}_B \cdot \nabla \Phi \\
  \end{array}
\right)\left(
         \begin{array}{c}
           {\bf u}^{(0)} \\
           \theta^{(0)} \\
         \end{array}
       \right)
=- \frac{\nabla \Phi}{\rho}\left(
     \begin{array}{c}
       p^{(0)} \\
       0 \\
     \end{array}
   \right),&
\ea
\ba{first1}
&\epsilon^{0}:~~~\, i\left(
  \begin{array}{cc}
    \frac{\partial \Phi}{\partial t} +{\bf u}_B \cdot \nabla \Phi & 0 \\
    0 & \frac{\partial \Phi}{\partial t} +{\bf u}_B \cdot \nabla \Phi \\
  \end{array}
\right)
\left(
  \begin{array}{c}
    {\bf u}^{(1)} \\
    \theta^{(1)} \\
  \end{array}
\right)
=-i \frac{\nabla \Phi}{\rho}
\left(
\begin{array}{c}
p^{(1)} \\
0 \\
\end{array}
\right)&
\nn\\
&-\left(
  \begin{array}{cc}
    \frac{\partial }{\partial t}  +{\bf u}_B \cdot \nabla + \mathcal{U}+ \widetilde \nu (\nabla \Phi)^2 & 0 \\
    0 & \frac{\partial }{\partial t}  +{\bf u}_B \cdot \nabla + \widetilde \kappa (\nabla \Phi)^2 \\
  \end{array}
\right)\left(
         \begin{array}{c}
           {\bf u}^{(0)} \\
           \theta^{(0)} \\
         \end{array}
       \right)&\nn\\
       &-\left(
  \begin{array}{cc}
    2\Omega\alpha\theta_B {\bf e}_r {\bf e}_{\varphi}^T & \alpha r \Omega^2 {\bf e}_r \\
    (\nabla \theta_B)^T & 0 \\
  \end{array}
\right)\left(
         \begin{array}{c}
           {\bf u}^{(0)} \\
           \theta^{(0)} \\
         \end{array}
       \right)-\frac{\nabla}{\rho}\left(
                                \begin{array}{c}
                                   p^{(0)} \\
                                  0 \\
                                \end{array}
                              \right).&
\ea
The amplitudes with the superscript $(0)$ are contained both in the equations \rf{first} corresponding to the lowest degree of $\epsilon$ equal to $-1$ and in the equations \rf{first1} corresponding to the degree $0$. Therefore, finding the lowest-order amplitudes with the superscript $(0)$ requires an iterative procedure involving both \rf{first} and \rf{first1}.

Taking the dot product of the first of the equations in the system \rf{first} with $\nabla \Phi$ under the constraint \rf{sc} we find that for $\nabla \Phi \ne 0$
\be{press}
p^{(0)}=0.
\ee

Under the condition \rf{press} the system  \rf{first} has a nontrivial solution if the determinant of the $4\times4$ matrix in its left-hand side is vanishing. This gives us a 4-fold characteristic root corresponding to the Hamilton-Jacobi equation
\be{hje}
\frac{\partial \Phi}{\partial t}  +{\bf u}_B \cdot \nabla \Phi=0
\ee
with the initial data: $\Phi({\bf x},0)=\Phi_0({\bf x})$.

Taking the gradient of \rf{hje} yields the eikonal equation \citep{LH1991}
\be{ghje}
\left(\frac{\partial }{\partial t}  +{\bf u}_B \cdot \nabla\right) \nabla\Phi+\nabla {\bf u}_B \cdot\nabla \Phi=0
\ee
with the initial condition $\nabla\Phi({\bf x},0)=\nabla\Phi_0({\bf x})$. In the notation ${\bf k}=\nabla \Phi$ and
$\frac{d}{d t}=\frac{\partial}{\partial t}+{\bf u}_B \cdot \nabla$
the eikonal equation \rf{ghje} is \citep{L1991}
\be{khje}
\frac{d {\bf k}}{d t}=-\nabla {\bf u}_B \cdot {\bf k}=-\mathcal{U}^T{\bf k},
\ee
where $\mathcal{U}^T$ denotes the transposed $3\times 3$ matrix $\mathcal{U}$  \citep{EY1995}.

Relations \rf{press} and \rf{hje} allow us to reduce the system  \rf{first1} to
\ba{second}
\left(\frac{d}{d t} +\mathcal{U}+ \widetilde \nu (\nabla \Phi)^2+2\alpha\Omega \theta_B {\bf e}_r {\bf e}_{\varphi}^T\right){\bf u}^{(0)}+\alpha r \Omega^2 {\bf e}_r \theta^{(0)}&=&-\frac{i\nabla \Phi}{\rho} p^{(1)},\\
(\nabla \theta_B)^T{\bf u}^{(0)}+\left(\frac{d}{d t} +\widetilde \kappa (\nabla \Phi)^2\right)\theta^{(0)}&=&0.
\ea
Multiplying the equation \rf{second} with $\nabla \Phi$ from the left, we isolate the pressure term
\be{p1}
p^{(1)}=i\rho \frac{\nabla \Phi}{(\nabla \Phi)^2}\cdot \left(\left[\frac{d}{d t} +\mathcal{U}+2\alpha\Omega \theta_B {\bf e}_r {\bf e}_{\varphi}^T\right]{\bf u}^{(0)}+\alpha r \Omega^2 {\bf e}_r \theta^{(0)} \right).
\ee

Taking into account the identity \citep{LH1991,KSF2014}
\be{iden}
\frac{d}{d t}(\nabla \Phi \cdot {\bf u}^{(0)})=\frac{d \nabla \Phi}{d t} \cdot {\bf u}^{(0)}+\nabla \Phi \cdot \frac{d {\bf u}^{(0)}}{d t}=0
\ee
we modify \rf{p1} in the following way
\be{p1m}
p^{(1)}=i\rho \frac{\nabla \Phi}{(\nabla \Phi)^2}\cdot \left(( \mathcal{U}+2\alpha\Omega \theta_B {\bf e}_r {\bf e}_{\varphi}^T){\bf u}^{(0)}+\alpha r \Omega^2 {\bf e}_r \theta^{(0)} \right)-i\rho \frac{1}{(\nabla \Phi)^2}\frac{d \nabla \Phi}{d t} \cdot {\bf u}^{(0)}.
\ee

Using now \rf{khje} we re-write \rf{p1m} in terms of the wave vector $\bf k$
\be{p1m4}
p^{(1)}=2i\rho \alpha\Omega \theta_B \frac{{\bf k}^T{\bf e}_r {\bf e}_{\varphi}^T}{|{\bf k}|^2}{\bf u}^{(0)}+i\rho \alpha r \Omega^2 \frac{{\bf k}^T{\bf e}_r}{|{\bf k}|^2}  \theta^{(0)} +2i\rho \frac{{\bf k}^T\mathcal{U}}{|{\bf k}|^2}  {\bf u}^{(0)}.
\ee
Substituting \rf{p1m4} into \rf{second} we finally arrive at the transport equations for the leading-order amplitudes ${\bf u}^{(0)}$ and $\theta^{(0)}$ in the expansions \rf{go} and \rf{go2}:
\ba{third}
\frac{d {\bf u}^{(0)}}{dt} &=& - \widetilde \nu |{\bf k}|^2{\bf u}^{(0)}-\left(\mathcal{I}-2\frac{{\bf k}{\bf k}^T}{|{\bf k}|^2}\right)\mathcal{U}{\bf u}^{(0)}-2\alpha\Omega \theta_B \left(\mathcal{I}-\frac{{\bf k}{\bf k}^T}{|{\bf k}|^2}\right){\bf e}_r {\bf e}_{\varphi}^T{\bf u}^{(0)}\nn\\
&-&\alpha r \Omega^2 \left(\mathcal{I}-\frac{{\bf k}{\bf k}^T}{|{\bf k}|^2}\right){\bf e}_r \theta^{(0)},\label{third1}\\
\frac{d \theta^{(0)}}{dt}&=&-\widetilde \kappa |{\bf k}|^2\theta^{(0)}-(\nabla \theta_B)^T{\bf u}^{(0)}\label{third2}
\ea
with the initial data ${\bf u}^{(0)}({\bf x},0)={\bf u}_0^{(0)}({\bf x})$ and $\theta^{(0)}({\bf x},0)=\theta_0^{(0)}({\bf x})$.

Note that the leading order terms dominate the solution \rf{go} and \rf{go2} for a sufficiently long time, provided that $\epsilon$ is small enough \citep{L1991,LSB1996}, which reduces analysis of instabilities to the investigation of the growth rates of solutions of the transport equations \rf{third1} and \rf{third2}.

Let us consider a fluid element with the trajectory passing through a point ${\bf x}_0$ at the initial moment $t=0$ :
\be{te}
\frac{d {\bf x}}{d t}={\bf u}_B, \quad {\bf x}(0)={\bf x}_0.
\ee

Then, the eikonal equation \rf{khje} can be interpreted as an ordinary differential equation describing the evolution of the wave vector ${\bf k}(t)=\nabla \Phi ({\bf x}(t),t)$ along the stream line \rf{te} with the initial condition ${\bf k}(0)={\bf k}_0=\nabla \Phi_0 ({\bf x}_0)$. Consequently, the transport equations \rf{third1} and \rf{third2} can also be treated as the system of ODEs along the stream lines of the flow for the
amplitudes ${\bf u}^{(0)}({\bf x}(t),t)$ and $\theta^{(0)}({\bf x}(t),t)$ with the initial data ${\bf u}^{(0)}(0)={\bf u}_0^{(0)}({\bf x}_0)$ and $\theta^{(0)}(0)=\theta_0^{(0)}({\bf x}_0)$.
Therefore, the characteristic equations \rf{te}, \rf{khje}, \rf{third1}, and \rf{third2} describe motion of the envelope of a perturbation localized at the initial moment of time at ${\bf x}_0$  along with the particles in the fluid flow that pass through ${\bf x}_0$ at $t=0$.

\section{Dispersion relation and its parameterizations}
Following the procedure described in \citep{FV1995} we
write the eikonal equation \rf{khje} in cylindrical coordinates $r,\varphi,z$  
\be{kvec}
\frac{\partial{\bf k}}{\partial t}
=\left(
                                  \begin{array}{ccc}
                                    0 & -2\Omega{\rm Ro} & 0 \\
                                    0 & 0 & 0 \\
                                    0 & 0 & 0 \\
                                  \end{array}
                                \right){\bf k}(t)
\ee
and consider the bounded, and asymptotically non-decaying and non-diverging solution of \rf{kvec} with $k_r = const $, $k_{\varphi}=0$, and $k_z = const$ \citep{FV1995,EY1995}. With this, we can write the amplitude transport
equations \rf{third} in the coordinate form as
\ba{coord}
\left[\frac{\partial }{\partial t}+\Omega \frac{\partial }{\partial{\varphi}} +\widetilde\nu|{\bf k}|^2\right] u_r^{(0)}
&=&\beta^{2}\left[2\Omega(1- \alpha\theta_B) u_{\varphi}^{(0)}-\alpha r \Omega^2\theta^{(0)}\right],\\
\left[\frac{\partial }{\partial t}+\Omega \frac{\partial }{\partial{\varphi}} +\widetilde\nu|{\bf k}|^2\right]u_{\varphi}^{(0)}&=&-2\Omega(1+{\rm Ro})u_r^{(0)},\label{coord2}\\
\left[\frac{\partial }{\partial t}+\Omega \frac{\partial }{\partial{\varphi}} +\widetilde \kappa |{\bf k}|^2\right]\theta^{(0)}&=&-D\Theta u_r^{(0)},\label{coord3}\\
\left[\frac{\partial }{\partial t}+\Omega \frac{\partial }{\partial{\varphi}} +\widetilde\nu|{\bf k}|^2\right]u_z^{(0)}
&=&-\beta^{2} \frac{k_r }{ k_z}\left[2\Omega (1-\alpha\theta_B)u_{\varphi}^{(0)} -\alpha r\Omega^2\theta^{(0)}\right],\label{coord4}
\ea
where $\beta=k_z|{\bf k}|^{-1}$ and $|{\bf k}|=\sqrt{k_{r}^{2}+k_{z}^{2}}$.

We see that the equations \rf{coord}, \rf{coord2}, and \rf{coord3} form a closed sub-system with respect to $u_r^{(0)}$, $u_{\varphi}^{(0)}$, and $\theta^{(0)}$. Indeed, the transformation $u_z^{(0)}\rightarrow (-k_r/k_z) u_r^{(0)} $ makes the equation \rf{coord4} equivalent to \rf{coord}. Following \citep{FV1995} we will look for a solution to this sub-system in the modal form as $\theta^{(0)}=\hat{\theta} e^{\lambda t +im\varphi}$ ,
and write the amplitude
equations in the matrix form $H\xi=\lambda \xi$, where $\xi=(\hat u_r,\hat u_{\varphi},\hat \theta)^T$ is the eigenvector associated with the eigenvalue $\lambda$ of the matrix\footnote{Cf. Eq.~\rf{ecomatrix} that is obtained in the narrow gap approximation.}
\ba{matrix}
H=\left(
  \begin{array}{ccc}
     -im\Omega -\widetilde\nu|{\bf k}|^2 & 2\Omega\beta^2(1-\alpha \Theta) & -\alpha r \Omega^2\beta^2 \\
    -2\Omega(1+{\rm Ro}) &  -im\Omega  -\widetilde \nu|{\bf k}|^2 & 0 \\
    -D\Theta & 0 &  -i m\Omega   -\widetilde \kappa |{\bf k}|^2 \\
  \end{array}
\right).
\ea
The solvability condition for a non-trivial solution yields the dispersion relation
\be{disp}
p(\lambda)=\det(H-\lambda I)=0,
\ee
where $I$ is the $3 \times 3$-unit matrix\footnote{Note that a similar dispersion relation was derived also in \cite{EM1980}  in the case when the radial acceleration $\bf g$ does not depend on $r$ (see Appendix). In our case ${\bf g}(r)={\bf g}_{cB}=(r\Omega^2,0,0)^T$.}.

Let us introduce the nondimensional parameters
\be{ndim}
 {\rm Ta}=\frac{\beta\Omega}{\widetilde \nu |{\bf k}|^2},\quad \hat{\Theta}=\frac{\Theta}{\Delta T}, \quad{ \gamma_a} = \alpha \Delta T, \quad{\rm Rt}=\frac{rD\Theta}{2\Theta}, \quad n=\frac{m}{\beta},\quad s =\frac{\lambda}{\beta \Omega},
\ee
where  ${\rm Ta}$ is the Taylor number, ${\rm Rt}$ is the thermal analog of the
Rossby number and it measures the local slope of the temperature
profile\footnote{not to be confused with the thermal Rossby number used in Atmospheric Physics \citep{Lappa}.}, $\Delta T$ is the temperature difference imposed at the cylindrical surfaces bounding the flow, $ \hat{\Theta}$ represents the dimensionless temperature, $n$ is a modified azimuthal wavenumber, and $s$ is the dimensionless eigenvalue. Introducing the diagonal matrix $R={\rm diag}(1,1,\Delta T/r)$, we transform \rf{matrix} into
\be{matrixndim}
R^{-1}HR=\beta\Omega\left(
  \begin{array}{ccc}
    -i n -\frac{1 }{\rm Ta} & 2 \beta(1-\gamma_a\hat\Theta) & -\gamma_a{\beta \Omega}  \\
    -\frac{2}{\beta}(1+{\rm Ro}) & -i n -\frac{1}{\rm Ta} & 0\\
    -2 \hat \Theta{\rm Rt}\frac{1 }{ \beta\Omega}  & 0 & -i  n -\frac{1 }{{\rm Ta} {\rm Pr}} \\
  \end{array}
\right)
\ee
and re-write the dispersion relation (\ref{disp}) in the equivalent form
\be{dr}
\det(R^{-1}HR- s \beta \Omega  I)=-\beta^3\Omega^3q(s)= 0,
\ee
where $q(s)$ is a third degree polynomial with complex coefficients that do not contain the parameters $\beta$ and $\Omega$ and depend entirely on the dimensionless control parameters ${\rm Ta}$, ${\rm Pr}$, $\hat{\Theta}$, $\gamma_a$, $n$ and on the logarithmic derivatives of the velocity and temperature profiles of the base flow represented by the dimensionless hydrodynamic and thermal Rossby numbers $\rm Ro$ and $\rm Rt$, respectively.

Introducing the new spectral variable $\chi=s+in$ into the complex dispersion relation $q(s)=0$, we transform it into a real third degree polynomial equation in $\chi$:
 \be{expl}
  \chi^3 + a_2\chi^2+a_1\chi + a_0 = 0.
 \ee
The real coefficients $a_i$ of the characteristic polynomial \rf{expl} depend on the dimensionless flow parameters as follows:
 \ba{coefdr}
 a_0 &=& \frac{1}{\rm Ta}\left[\frac{1}{{\rm Pr}{\rm Ta}^2}+\frac{4(1+{\rm Ro})(1-\gamma_a \hat{\Theta})}{{\rm Pr}}      -2\gamma_a \hat{\Theta} {\rm Rt}\right],  \\
 a_1&=& 4(1+{\rm Ro})(1-\gamma_a \hat{\Theta})-2\gamma_a \hat{\Theta}
 {\rm Rt}+\frac{{\rm Pr}+2}{{\rm Pr}{\rm Ta}^2}, \\
 a_2&=& \frac{2{\rm Pr}+1}{{\rm Ta} {\rm Pr}}.
 \ea

\begin{table}
\begin{center}
\begin{tabular}{lcrl}

 Rotation case  &\quad ${\rm Ro}$ &\quad  ${\rm Pr}^{S}$ &\quad  ${\rm Pr}^{H}$ \\

Inner cylinder &\quad  $- \frac{1}{1-\eta}$ &\quad  $-\frac{4\eta {\rm ln}\eta}{1-\eta}(\frac{1}{\gamma_a}-\frac{1}{2})$ &\quad
$-\left[1+\frac{8\eta {\rm ln}\, \eta}{1-\eta}(\frac{1}{\gamma_a}-\frac{1}{2})\right]^{-1}$ \\

Outer cylinder &\quad  $\frac{\eta}{1-\eta}$ &\quad
$\frac{4{\rm ln}\,\eta}{1-\eta}(\frac{1}{\gamma_a}-\frac{1}{2})$ &\quad
$-\left[1-\frac{8{\rm ln}\, \eta}{1-\eta}(\frac{1}{\gamma_a}-\frac{1}{2})\right]^{-1}$\\

Keplerian &\quad  $-\frac{1+\sqrt{\eta}+\eta}{(1+\sqrt{\eta})^2}$ &\quad
$\frac{4\sqrt{\eta}\,{\rm ln}\,\eta}{(1+\sqrt{\eta})^2} (\frac{1}{\gamma_a}-\frac{1}{2})$ &\quad
$-\left[1-\frac{8\sqrt{\eta}{\rm ln}\, \eta}{(1+\sqrt{\eta})^2}(\frac{1}{\gamma_a}-\frac{1}{2}) \right]^{-1}$  \\

Solid body  &\quad $0$&\quad  $4{\rm ln}\,\eta(\frac{1}{\gamma_a}-\frac{1}{2})$&\quad  $-\left[1-8{\rm ln}\,\eta(\frac{1}{\gamma_a}-\frac{1}{2})\right]^{-1}$\\

\end{tabular}
\caption {Rossby numbers for different rotating cases of the Couette-Taylor system according to Eq.~\rf{meanpar} and the corresponding asymptotic values \rf{TaSAOAGamma} that bound the intervals of $\rm Pr$ at which stationary (${\rm Pr}^{S}$) and oscillatory (${\rm Pr}^{H}$) modes are admissible. }
\end{center}
\label{tab1}
\end{table}

The set of dimensionless parameters \rf{ndim} is convenient for applications in astrophysical gasdynamics \citep{AG1978,BP2016}. In order to facilitate comparison with the experimental data we need to express these parameters in terms of the realistic Taylor-Couette system. In this case, the base flow state confined in the gap between two coaxial cylinders of radii $a$ and $b$ rotating with the angular frequencies $\Omega_a$ and $\Omega_b$, respectively,  is given by \citep{Chandrasekhar, MYM2015, Ali}
\be{BF}
\Omega=\frac{\mu-\eta^2}{1-\eta^2}+\frac{1-\mu}{1-\eta^2}\frac{\eta^2}{(1-\eta)^2}\frac{1}{r^2},  \quad \hat{\Theta} = \frac{{\rm ln}(1-\eta)r}{{\rm ln}\eta},
\ee
where  $\eta=a/b < 1$ is the radius ratio and $\mu=\Omega_b/\Omega_a$ is the rotation ratio. The angular velocity is measured in the units of $\Omega_a$ and the radial distance in the units of $d = b-a $.

Evaluating the parameters \rf{ro}, \rf{ndim}, and \rf{BF} at the geometric mean radius  $r_g = \sqrt{\eta}/(1-\eta)$ (as proposed by \cite{Dubrulle}), we find\footnote{In the case of counter-rotating cylinders, the hydrodynamic Rossby number diverges at $\mu = -\eta$. For $\mu \le -\eta$, one may choose the geometric mean radius of the potentially unstable zone.} 
\be{meanpar}
\hat{\Theta}=\frac{1}{2}, \quad {\rm Rt} =
\frac{1}{{\rm ln} \eta}, \quad {\rm Ro}= -1+\frac{\mu-\eta^2}{(1-\eta)(\eta+\mu)}.
\ee

Relations \rf{meanpar} allow us to interpret the local Rossby numbers $\rm Ro$ and $\rm Rt$ in terms of the
radius and velocity ratios $\eta$ and $\mu$ of the realistic Taylor-Couette cell, see Table~1 containing the
four different rotation cases: The sole rotation of the inner cylinder ($\mu =0$), the Keplerian rotation ($\mu = \eta^{3/2}$), the sole rotation of the outer cylinder ($\mu \rightarrow \infty$), and the solid body rotation $(\mu=1)$. In particular,
${\rm Rt}<0$ for the Taylor-Couette system due to the geometric constraint $0< \eta < 1$ and the temperature variation is such that $0\le \hat{\Theta}\le 1 $.

\section{Local instabilities}
 \subsection{Diffusionless instabilities}
In the case when the viscosity and thermal diffusivity are both set to zero in \rf{matrix} or, equivalently, ${\rm Ta} \rightarrow \infty$ in \rf{matrixndim}, the dispersion relation \rf{expl} factorizes to
\be{difless}
\chi\left(\chi^2+4(1+{\rm Ro})(1-\gamma_a \hat{\Theta})-2\gamma_a \hat{\Theta}
 {\rm Rt}\right)=0.
\ee

The root $\chi=0$ corresponds  to the pure imaginary eigenvalue $\lambda=\beta\Omega s=-im\Omega$ describing a stable rotating wave with the local frequency $\Omega$. Similarly, the flow admits stable inertial waves with frequencies given by
\be{freq}
\frac{\omega}{\Omega} = -m\pm\beta\sqrt{4(1+{\rm Ro})(1-\gamma_a \hat{\Theta})-2 \gamma_a \hat{\Theta} {\rm Rt}},
\ee
if the radicand in \rf{freq} is positive:
 \be{rcr}
4(1+{\rm Ro})(1-\gamma_a\hat{\Theta})-2\gamma_a\hat{\Theta}{\rm Rt}> 0.
\ee
These inertial waves result from the combined effects of the rotation and temperature gradient.
While the term $\gamma_a \hat{\Theta}$ is very weak in the Boussinesq approximation, the term $\gamma_a \hat{\Theta} {\rm Rt}$ can become significant depending on the steepness of the temperature profile.
In the isothermal case ($ \gamma_a=0$), we retrieve the frequency of the Kelvin
waves in rotating flows.

The inequality \rf{rcr} contains the generalized Rayleigh criterion derived in \citep{MYM2015} and \citep{MUBA2002} for the stability of the inviscid flow with curved streamlines and a radial temperature gradient.
Indeed,  \cite{MUBA2002} and \cite{MYM2015} found the criterion for the stability against axisymmetric perturbations in the diffusionless case in terms of the generalized Rayleigh discriminant $\Psi(r)$:
\be{grc}
\Psi(r)>0, \quad \Psi(r)=\Phi-\gamma_a\left(\hat{\Theta} \Phi+\frac{d \hat{\Theta}}{d r}\frac{V^2}{r} \right), \quad \Phi(r)=\frac{1}{r^3}\frac{d(rV)^2}{dr}.
\ee

Expressing the classical hydrodynamic Rayleigh discriminant $\Phi(r)$ via the hydrodynamic Rossby number as
$
\Phi=4\Omega^2(1+{\rm Ro})
$, we obtain
\be{psiro}
\Psi=\Omega^2\left\lbrace 4(1+{\rm Ro})-2\gamma_a\hat{\Theta}\left[{\rm Rt}+2(1+{\rm Ro})\right] \right\rbrace.
\ee

Requiring $\Psi>0$ we arrive exactly at the criterion \rf{rcr}. The opposite inequality $\Psi < 0$ or $4(1+{\rm Ro})(1-\gamma_a\hat{\Theta})-2\gamma_a\hat{\Theta}{\rm Rt} < 0$ coincides with the condition for the instability to stationary axisymmetric perturbations known as the \textit{diffusionless Goldreich-Schubert-Fricke (GSF) instability} \citep{AG1978}.
%
\subsection{Enhancement of stability by viscosity and thermal diffusivity at ${\rm Pr}=1$ }
In the case when ${\rm Pr}=1$ (i.e. when the viscous and thermal diffusion timescales are equal) the dispersion relation \rf{expl} can also be factorized:
\be{pr1}
\left(\chi+\frac{1}{{\rm Ta}}\right)\left[\left(\chi+\frac{1}{{\rm Ta}}\right)^2+4(1-\gamma_a\hat\Theta)(1+{\rm Ro})-2\gamma_a\hat\Theta{\rm Rt}\right]=0.
\ee
By this reason, the roots of the dispersion relation \rf{pr1} can be found explicitly:
\ba{eig1}
s_{1,2}&=&-\frac{1}{{\rm Ta}}- i n \pm i \sqrt{4(1+{\rm Ro})(1-\gamma_a \hat{\Theta})-2\gamma_a\hat{\Theta}{\rm Rt}},\\
s_3&=&-\frac{1}{{\rm Ta}}- i n. \label{eig13}
\ea
The mode with the eigenvalue $s_3$ is a pure hydrodynamic mode as it does not contain the temperature gradient, it is damped  by the viscosity. The modes corresponding to the eigenvalue
$s_{1,2}$ contain both the hydrodynamic and thermal effects. One of the modes is always damped, the other one should be excited when its growth rate vanishes, i.e. when
\be{marg}
4(1+{\rm Ro})(1-\gamma_a \hat{\Theta})- 2 \gamma_a \hat{\Theta}{\rm Rt}+\frac{1}{{\rm Ta}^2}=0.
\ee

The double-diffusive stability criterion given by the requirement for the left hand side of Eq.~\rf{marg} to be positive suggests that at $\rm Pr=1$ the viscosity and thermal diffusivity enlarge the stability domain of the diffusionless system.
Furthermore, in the limit ${\rm Ta} \rightarrow \infty$ the roots \rf{eig1} and \rf{eig13} reduce to the roots of the diffusionless dispersion relation \rf{difless}, whereas the double-diffusive stability criterion reduces to the diffusionless criterion \rf{rcr}.

Note that the limiting procedure that first makes the diffusion coefficients of a dissipative system with two diffusion mechanisms equal and then tends them to zero typically yields a correct diffusionless stability criterion in many double-diffusive systems of hydrodynamics and magnetohydrodynamics \citep{KSF2014,K2016}. In general, the limit of zero dissipation of the double-diffusive stability criteria should not necessarily coincide with the diffusionless stability criteria \citep{AG1978,KV2010,K2013}.

\subsection{Double-diffusive flow stability criteria at arbitrary $\rm Pr$}
We derive stability conditions of the base flow at arbitrary $\rm Pr$ by applying the Lienard-Chipart stability criterion \citep{K2013} to the real polynomial \rf{expl} of degree 3 in $\chi$:
\be{ls}
a_0>0, \quad a_2>0,\quad -a_0(a_0-a_1a_2)>0.
\ee

The second inequality is always fulfilled whereas the first one yields:
\be{bsc1}
4(1+{\rm Ro})(1-\gamma_a\hat{\Theta})-2\gamma_a\hat{\Theta}{\rm Rt} {\rm Pr} + \frac{1}{{\rm Ta}^2} >0.
\ee
The condition \rf{bsc1} is a modified Rayleigh criterion that takes into account both the kinematic viscosity and the thermal diffusion. Writing it as
$$
1+{\rm Ro}>\frac{2\gamma_a\hat{\Theta} \,{\rm Rt}{\rm Pr}}{4(1-\gamma_a\hat{\Theta})}-\frac{1}{4{\rm Ta}^2(1-\gamma_a\hat{\Theta})},
$$
we conclude that, for ${\rm Rt}<0$ and $0\le \hat{\Theta} \le 1$,  the kinematic viscosity and the outward heating $(\gamma_a >0)$ are stabilizing and the inward heating $(\gamma_a < 0)$  is destabilizing with respect to the Rayleigh criterion ($1+{\rm Ro}>0$) for the ideal incompressible fluid.

At ${\rm Pr}=1$ the condition \rf{bsc1} reduces to the criterion \rf{marg}. Furthermore, the limit ${\rm Ta}\rightarrow \infty$ applied to \rf{bsc1} yields the diffusionless criterion \rf{rcr} only if ${\rm Pr}=1$.

The condition $a_0= 0$ yields the control parameter ${\rm Ta}$ as a function of the other control parameters:
\be{stmodes}
{\rm Ta}_c^{S} = \frac{1}{\sqrt{2\gamma_a \hat{\Theta} {\rm Rt}{\rm Pr}-4(1+{\rm Ro})(1-\gamma_a \hat{\Theta})}}.
\ee
Since at $a_0= 0$ the dispersion equation \rf{expl} has the root $\chi=0$, corresponding to $s=-in$, the relation \rf{stmodes} is the threshold of the \textit{double-diffusive} Goldreich-Schubert-Fricke instability \citep{AG1978}, which is a stationary axisymmetric instability in the case when $n=0$.

The third inequality of \rf{ls} yields
 \ba{4x4}
-\left(\gamma_a\hat{\Theta}({\rm Pr}+1){\rm Rt}-4{\rm Pr}(1-\gamma_a\hat{\Theta})(1+{\rm Ro})-\frac{({\rm Pr}+1)^2}{{\rm Pr}{\rm Ta}^2}\right)>0,
\ea
from which the threshold of the oscillatory instability through a Hopf bifurcation naturally follows:
\be{Hopf}
{\rm Ta}_c^{H}= \frac{{\rm Pr}+1}{\sqrt{{\rm Pr} ({\rm Pr}+1)\gamma_a\hat{\Theta}{\rm Rt} -4 {\rm Pr}^2(1-\gamma_a\hat{\Theta})(1+{\rm Ro})}}.
\ee

Indeed, from the Vieta's formulas \citep{V2003}
 \be{rootsprop}
\chi_1+\chi_2+\chi_3 = -a_2,\quad \chi_1\chi_2+\chi_2\chi_3+\chi_3\chi_1 = a_1, \quad \chi_1\chi_2\chi_3 = -a_0.
\ee
it follows that at ${\rm Ta}={\rm Ta}_c^H$ the characteristic equation \rf{expl} admits two imaginary roots $\chi_{1,2}=\pm i\bar{\omega}_c$, corresponding to $s=-in\pm i\bar{\omega}_c$ where $n$ can  be either  zero or not, and one real root $\chi_3= -a_2$. The real root corresponds to a damped mode while the imaginary roots yield the Hopf frequency $\bar{\omega}_c$ given by
\be{imagroot}
\bar{\omega}_c= \frac{\sqrt{4(1+{\rm Ro})(1-\gamma_a\hat{\Theta})-\gamma_a\hat{\Theta}{\rm Rt}{\rm Pr}({\rm Pr}+1)}}{{\rm Pr}+1}
\ee
and describe a marginal oscillatory mode. The total frequency of this marginal mode is given by $\omega_c=-\beta\Omega(n\pm\bar{\omega}_c)$. The marginal value
${\rm Ta}_c^{H}$ corresponds to a Hopf bifurcation of the base flow to a state oscillating with a Hopf frequency $\bar{\omega}_c$. For $n=0$, the marginal mode is an oscillatory axisymmetric mode while for $n \neq 0$, the marginal mode is an oscillatory non-axisymmetric mode.

\subsection{Codimension-2 points in the case of the Rayleigh-unstable flows}
In the parameter space, the boundary of the stationary instability \rf{stmodes}, corresponding to a simple zero root, and the boundary of the Hopf bifurcation \rf{Hopf}, corresponding to a pair of imaginary roots, may have common codimension-2 points with the coordinates $({\rm Pr}^*, {\rm Ta}^*)$ given for ${\rm Ro}+1<0$ and $\gamma_a>0$ by
\ba{codim2a}
{\rm Pr}^{*}&=&-\frac{1}{2}+\frac{1}{2}\sqrt{1+16\frac{(1-\gamma_a\hat{\Theta})}{\gamma_a\hat{\Theta}}\frac{(1+{\rm Ro})}{ {\rm Rt}}},  \\
{\rm Ta}^{*}&=& \frac{1}{\sqrt{2\gamma_a\hat{\Theta}{\rm Rt}{\rm Pr}^{*}-4(1-\gamma_a\hat{\Theta})(1+{\rm Ro})}}. \label{codim2ab}
\ea

\subsection{The Rayleigh line}

In the particular case ${\rm Ro}=-1$, corresponding to the Rayleigh line, the roots of the dispersion relation are found explicitly for an arbitrary $\rm Pr$:
$$
\chi_{1,2}=-\frac{\rm Pr+1}{\rm 2TaPr}\pm\sqrt{2{\rm Rt}\hat\Theta\gamma_a+\frac{({\rm Pr}-1)^2}{4{\rm Ta^2}{\rm Pr}^2}},\quad\chi_3=-\frac{1}{\rm Ta}.
$$
One of the first two roots vanishes at the threshold of stationary instability ${\rm Ta}={\rm Ta}_c^{S}$, where $(\gamma_a<0, {\rm Rt}<0)$
$$
{\rm Ta}^{S}_c=\frac{1}{\sqrt{2\gamma_a\hat\Theta{\rm Rt}{\rm Pr}}}.
$$
Thus, at the Rayleigh line the stationary instability is possible for the inward heating only. When the heating is outward and ${\rm Rt}<0$,
both the criterion \rf{bsc1} and \rf{4x4} is fulfilled and the flow is stable.

\begin{table}
\begin{center}
\begin{tabular}{llll}

Rot.  &~ ${\rm Ta}_c^{S}$ &~  ${\rm Ta}_c^{H}$ &~  $\omega_c/\Omega$ \\

Inn. &~  $\left[\frac{2\left(2-{\gamma_a} \right)\eta}{1-\eta}{+}\frac{\gamma_a{\rm Pr} }{{\rm ln} \eta}  \right]^{-\frac{1}{2}}$ &~  $\frac{1+{\rm Pr}}{\rm Pr}\left[\frac{2(2-\gamma_a)\eta}{1-\eta} {+}\frac{\gamma_a }{2{\rm ln} \eta}\frac{1+\rm Pr}{\rm Pr} \right]^{-\frac{1}{2}}$ &~
$\frac{\beta}{1+\rm Pr}\left[\frac{2\eta(\gamma_a-2)}{1-\eta}{-}\frac{\gamma_a{\rm Pr}(1+{\rm Pr})}{2 {\rm ln}\eta}\right]^{\frac{1}{2}}$ \\

Ray. &~ $\left[\frac{\gamma_a{\rm Pr} }{{\rm ln} \eta} \right]^{-\frac{1}{2}}$ &~ &~ \\

Sol. &~ $\left[2(\gamma_a{-}2){+}\frac{\gamma_a{\rm Pr} }{{\rm ln} \eta} \right]^{-\frac{1}{2}}$&~  $\frac{1+{\rm Pr}}{\rm Pr}\left[2(\gamma_a{-}2) {+}\frac{\gamma_a }{2{\rm ln} \eta}\frac{1+\rm Pr}{\rm Pr} \right]^{-\frac{1}{2}}$ &~  $\frac{\beta}{1+\rm Pr}\left[{2(2{-}\gamma_a)}{-}\frac{\gamma_a{\rm Pr}(1+{\rm Pr})}{2 {\rm ln}\eta}\right]^{\frac{1}{2}}$\\

Kep. &~  $\left[\frac{2\sqrt{\eta}\left(2-{\gamma_a}\right)}{(1+\sqrt{\eta})^2}{+}\frac{\gamma_a{\rm Pr} }{{\rm ln} \eta}\right]^{-\frac{1}{2}}$ &~
$\frac{1+\rm Pr}{\rm Pr}\left[\frac{2\sqrt{\eta}(2-\gamma_a)}{(1+\sqrt{\eta})^2}{+}\frac{\gamma_a }{2{\rm ln} \eta}\frac{1+\rm Pr}{\rm Pr}\right]^{-\frac{1}{2}}$ &~
$\frac{\beta}{1+\rm Pr}\left[\frac{2\sqrt{\eta}(\gamma_a-2)}{(1+\sqrt{\eta})^2}{-}\frac{\gamma_a{\rm Pr}(1+{\rm Pr})}{2 {\rm ln}\eta}\right]^{\frac{1}{2}}$  \\

Out. &~  $\left[\frac{2\left(2-{\gamma_a}\right)}{\eta-1}{+}\frac{\gamma_a{\rm Pr} }{{\rm ln} \eta}\right]^{-\frac{1}{2}}$ &~
$\frac{1+\rm Pr}{\rm Pr}\left[\frac{2(\gamma_a-2)}{1-\eta} {+}\frac{\gamma_a }{2{\rm ln} \eta}\frac{1+\rm Pr}{\rm Pr}\right]^{-\frac{1}{2}}$ &~
$\frac{\beta}{1+\rm Pr}\left[\frac{2(2-\gamma_a)}{1-\eta}{-}\frac{\gamma_a{\rm Pr}(1+{\rm Pr})}{2 {\rm ln}\eta}\right]^{\frac{1}{2}}$\\

\end{tabular}
\caption {Thresholds of the critical modes and the critical Hopf frequency for different rotation regimes in the Couette-Taylor system: (Inn.) inner cylinder rotating, (Ray.) Rayleigh line, (Sol.) solid body rotation, (Kep.) Keplerian rotation, (Out.) outer cylinder rotating. }
\end{center}
\label{tab2}
\end{table}

\section{Stationary and oscillatory modes}

The relations \rf{stmodes} and \rf{Hopf} contain besides ${\rm Pr}$,  the group of parameters ${\rm Ro}$, $\gamma_a\hat{\Theta}$ and $\gamma_a\hat{\Theta}{\rm Rt}$, $\rm Rt <0$. Therefore, the conditions of the occurrence of the stationary or oscillatory modes can be analyzed in terms of these parameters.

Let us define the Brunt-V\"ais\"al\"a frequency $\rm N$ as \citep{MYM2015}\footnote{These authors have defined ${\rm N}^2$ with the opposite sign.}
$${\rm N}^2=\frac{1}{\rho_0}\nabla{\rho}\cdot{\bf g}_{cB} = \frac{1}{\rho_0} \frac{d\rho}{dr}\frac{V^2}{r}.$$
Taking into account that $\rho(r)=\rho_0(1-\alpha \theta(r))$ in the Boussinesq approximation, we conclude that
${\rm N}^2=-2\gamma_a\hat{\Theta} {\rm Rt}\,\Omega^2$.

In the diffusionless case, for ${\rm Ro} < -1$, the flow is Rayleigh unstable, i.e. the radial stratification of the square of the angular momentum is negative; for ${\rm Ro} > -1$, the flow is Rayleigh stable \citep{Chandrasekhar}.
The centrifugal acceleration is oriented away from  the rotation axis, implying that  the outward heating  ($\gamma_a > 0$) corresponds to a stable stratification of the density (${\rm N}^2 > 0$) while the inward heating ($\gamma_a < 0$) corresponds to an unstable density stratification, characterized by the negative sign of the squared Brunt-V\"ais\"al\"a frequency $({\rm N}^2<0)$.

For Rayleigh stable flow configurations, the negative stratification of the density yields a thermal instability analog to the Rayleigh-B\'{e}nard convection which might lead to stationary modes at the onset. For Rayleigh unstable configurations, the negative density stratification enforces the centrifugal instability
and decreases the threshold compared to the isothermal flow; while the positive density stratification will delay the threshold and inertial waves may be excited by the rotation when ${\rm N}^2$ becomes significant. These inertial waves, observed in numerical simulations, have been recently reported in \citep{MYM2015}.

\subsection{Stationary modes}
Taking into account the relations \rf{meanpar} we write the threshold for the onset of the stationary instability
\rf{stmodes} in terms of the parameters of the Couette-Taylor flow
\be{CTSA}
r_{S}=\frac{{\rm Ta}_c^{S}}{{\rm Ta}_0} = \left[1-\frac{\gamma_a}{2}+\frac{\gamma_{a}{\rm Pr}}{{\rm ln} \eta}\left({\rm Ta}_0\right)^2\right]^{-1/2},
\ee
where we have introduced the threshold for the isothermal flow as follows
\be{iso}
{\rm Ta}_c^{S}(\gamma_a=0) = \frac{1}{2}\sqrt{\frac{-1}{{\rm Ro}+1}} \equiv {\rm Ta}_0.
\ee
For the circular Couette flow, the quantity $({\rm Ta}_0)^2 $ is positive when  ${\rm Ro}+1<0$ (or $\mu < {\eta}^2$), i.e. in the Rayleigh unstable zone  and $({\rm Ta}_0)^2 < 0$ when ${\rm Ro}+1>0$, i.e. in the Rayleigh stable zone \citep{Chandrasekhar}. The explicit expressions for ${\rm Ta}_c^{S}$ calculated for different rotating regimes of the Couette-Taylor flow are given in Table~2.

\begin{figure}
    \begin{center}
    \includegraphics[angle=0, width=0.32\textwidth]{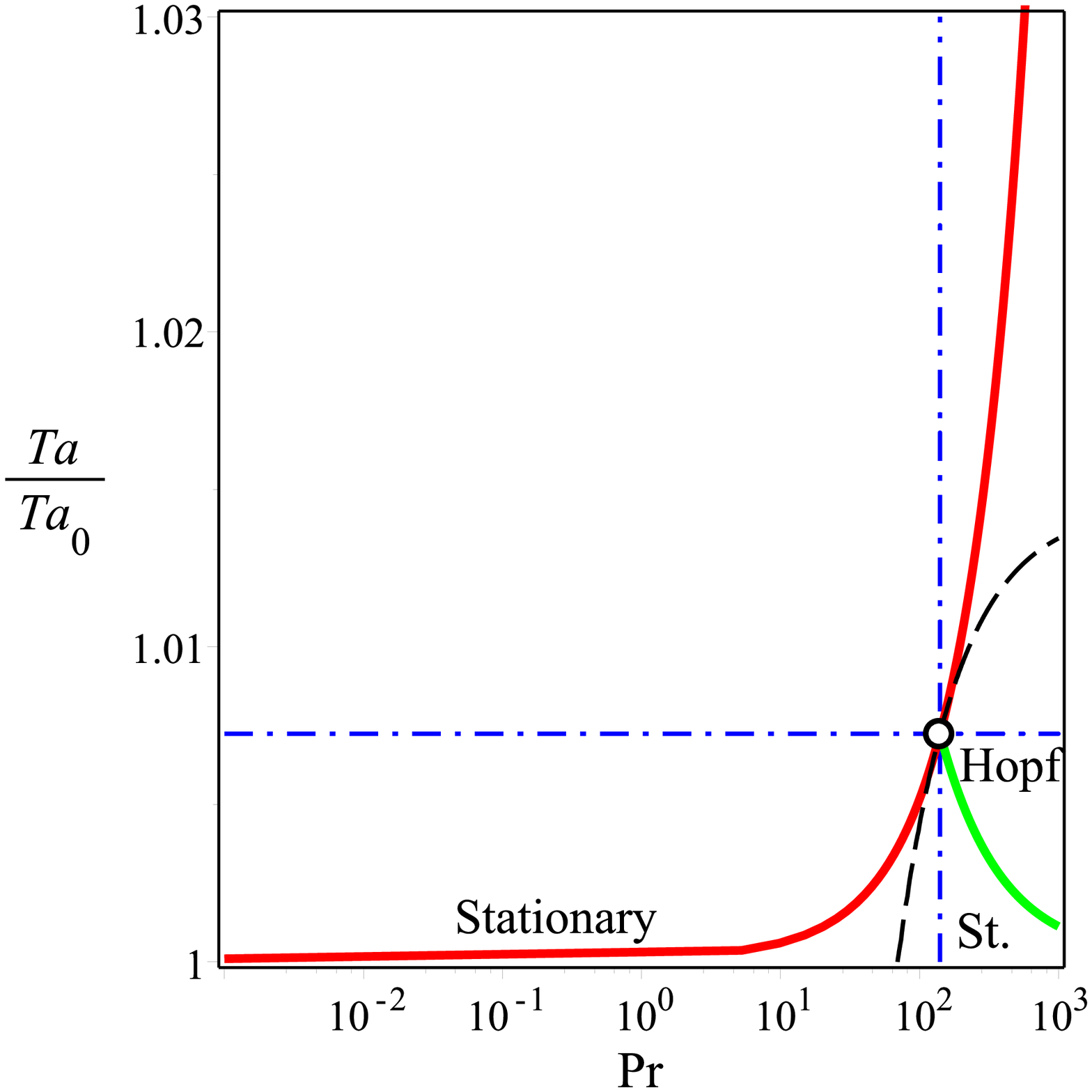}
    \includegraphics[angle=0, width=0.32\textwidth]{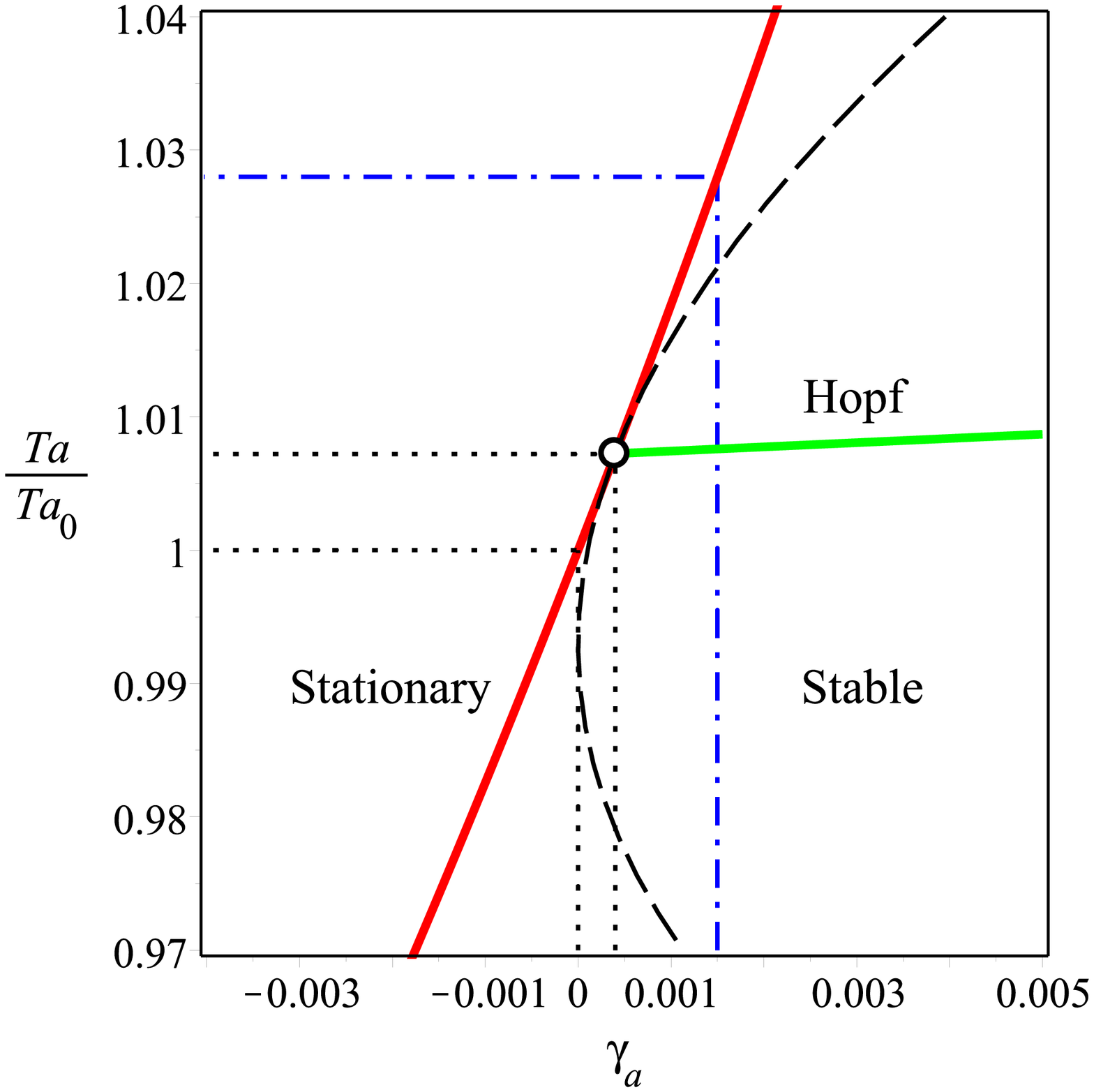}
    \includegraphics[angle=0, width=0.32\textwidth]{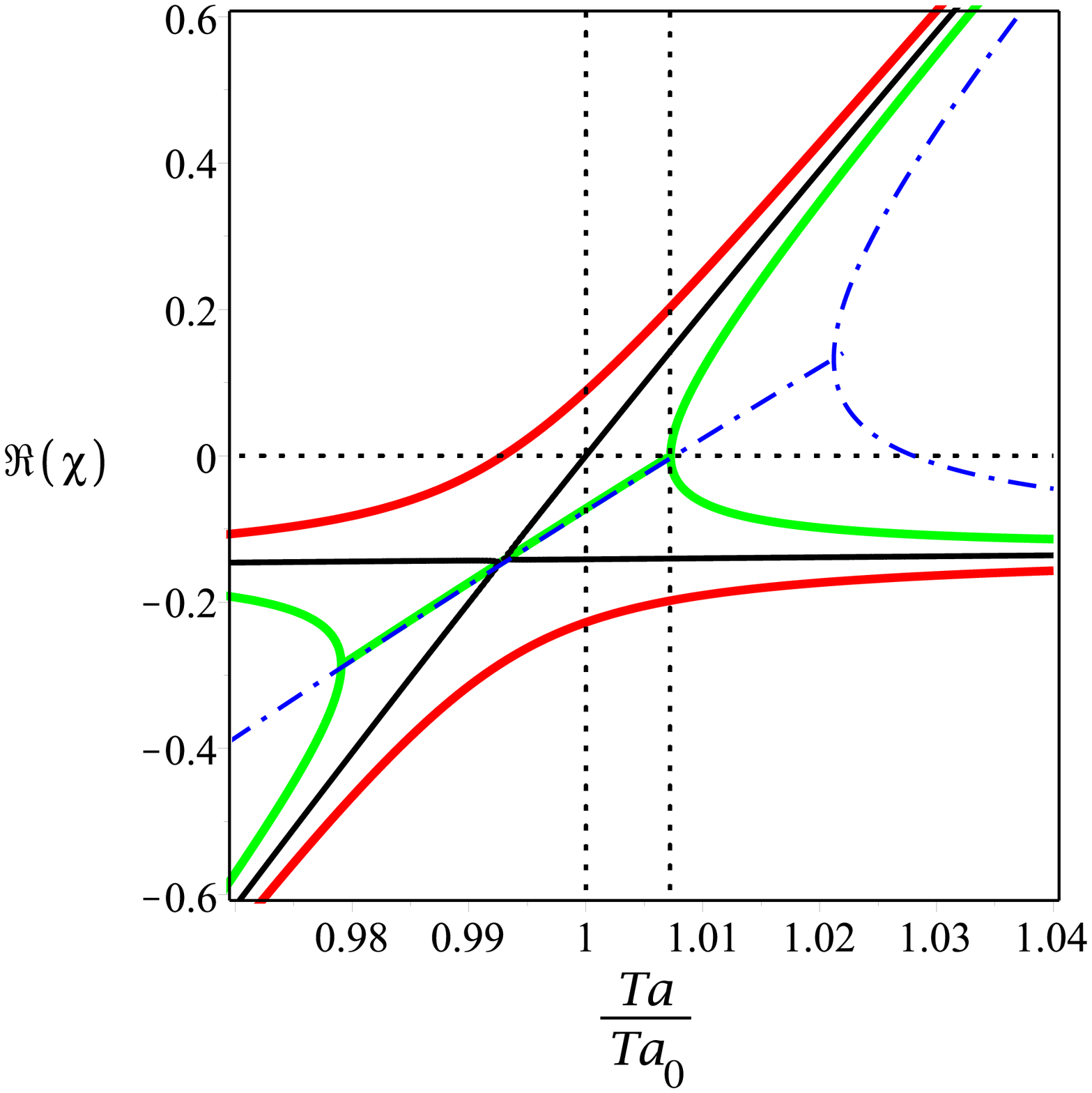}
    \end{center}
    \caption{ Rotating inner cylinder with $\mu=0$, $\eta=0.99$. (Left) Stability diagram at $\gamma_a=0.0004$ showing the boundaries of stationary instability \rf{CTSA} and Hopf bifurcation \rf{TaSAOAGamma} that have a common codimension-2 point $({\rm Pr^*}\approx140.55,{\rm Ta^*}/{\rm Ta_0}\approx1.0072)$ given by Eqs.~\rf{CD2pt} and \rf{CD2pta}.  The black dashed line corresponds to double real roots $\chi$: $\chi<0$ in the stability domain, $\chi>0$ in the instability domain, and $\chi=0$ at the codimension-2 point. (Center) The critical values of ${\rm Ta}/{\rm Ta}_0$ versus $\gamma_a$   at ${\rm Pr}\approx140.55$ with the codimension-2 point $(\gamma_a=0.0004, {\rm Ta}/{\rm Ta}_0\approx1.0072$) and the (black dashed) line of real eigenvalues. (Right) Growth rates (${\rm Re}\chi$) at ${\rm Pr}\approx140.55$ and (red) $\gamma_a=-0.0004$, (black) $\gamma_a=0$, (green) $\gamma_a=0.0004$, (dashdot blue) $\gamma_a=0.0015$. }
    \label{fig1}
    \end{figure}

In Figure~\ref{fig1} we plot the threshold \rf{CTSA} in the Rayleigh-unstable case of the inner cylinder rotation with $\mu=0$ and $\rm Ro+1=-\eta/(1-\eta)<0$, see Table~1. We see that the ratio $r_{S} < 1$, i.e. there is a destabilization, for the inward heating $(\gamma_a<0)$,  and $r_{S} > 1$ for the outward heating $(\gamma_a>0)$.
Note that the threshold of the stationary modes \rf{CTSA} is a function of the parameter $\widehat{S}= -\gamma_a{\rm Pr}/{\rm ln} \eta$, where $|\gamma_a|\ll 1$ in the Boussinesq approximation. This result theoretically justifies the numerical analysis of \citep{MYM2015} where it was shown that the stationary critical modes were almost scaled by the temperature drop coefficient $S= \widehat{S}\eta/(1-\eta)$.

                   \begin{figure}
    \begin{center}
    \includegraphics[angle=0, width=0.41\textwidth]{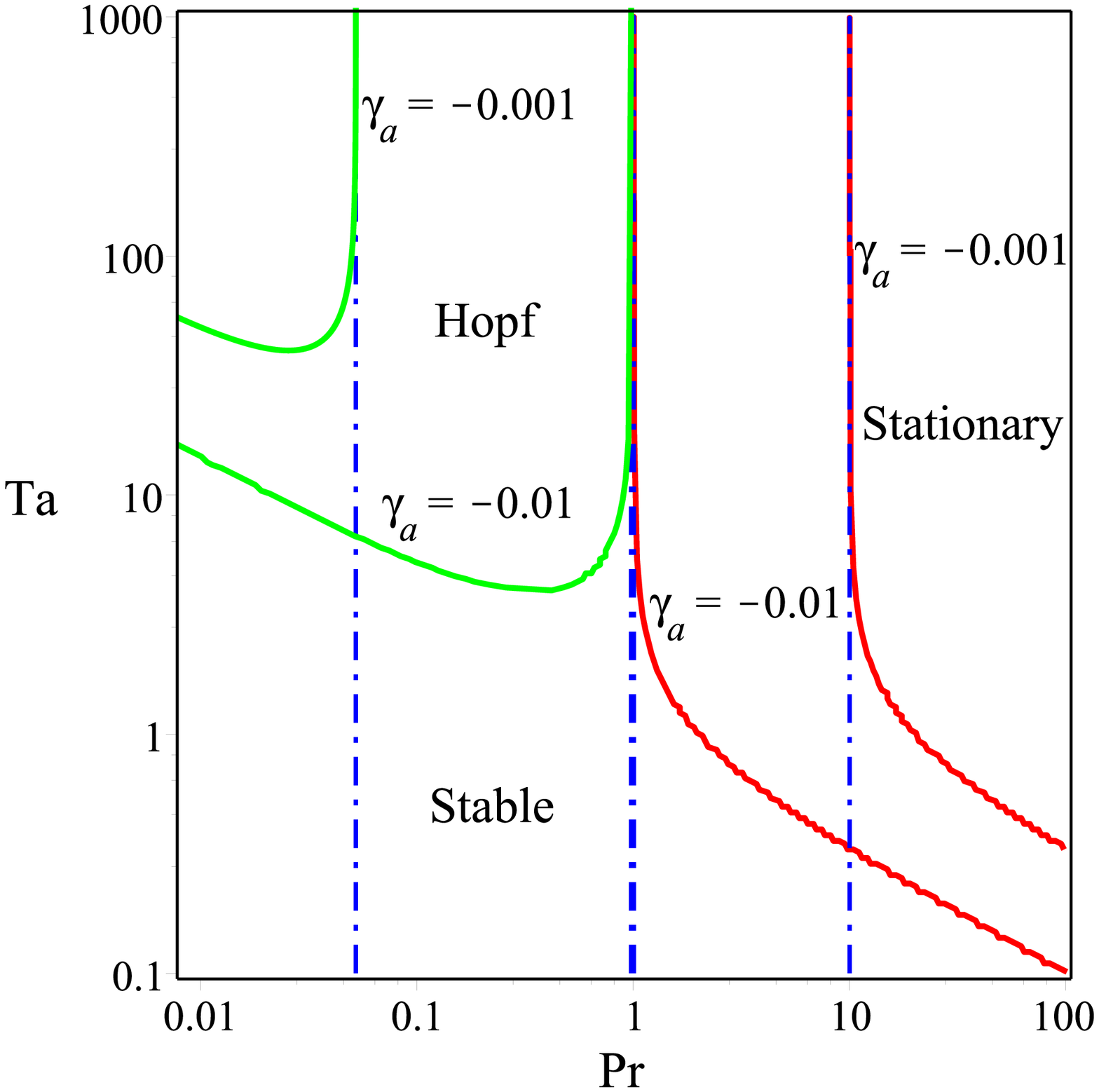}
    \includegraphics[angle=0, width=0.41\textwidth]{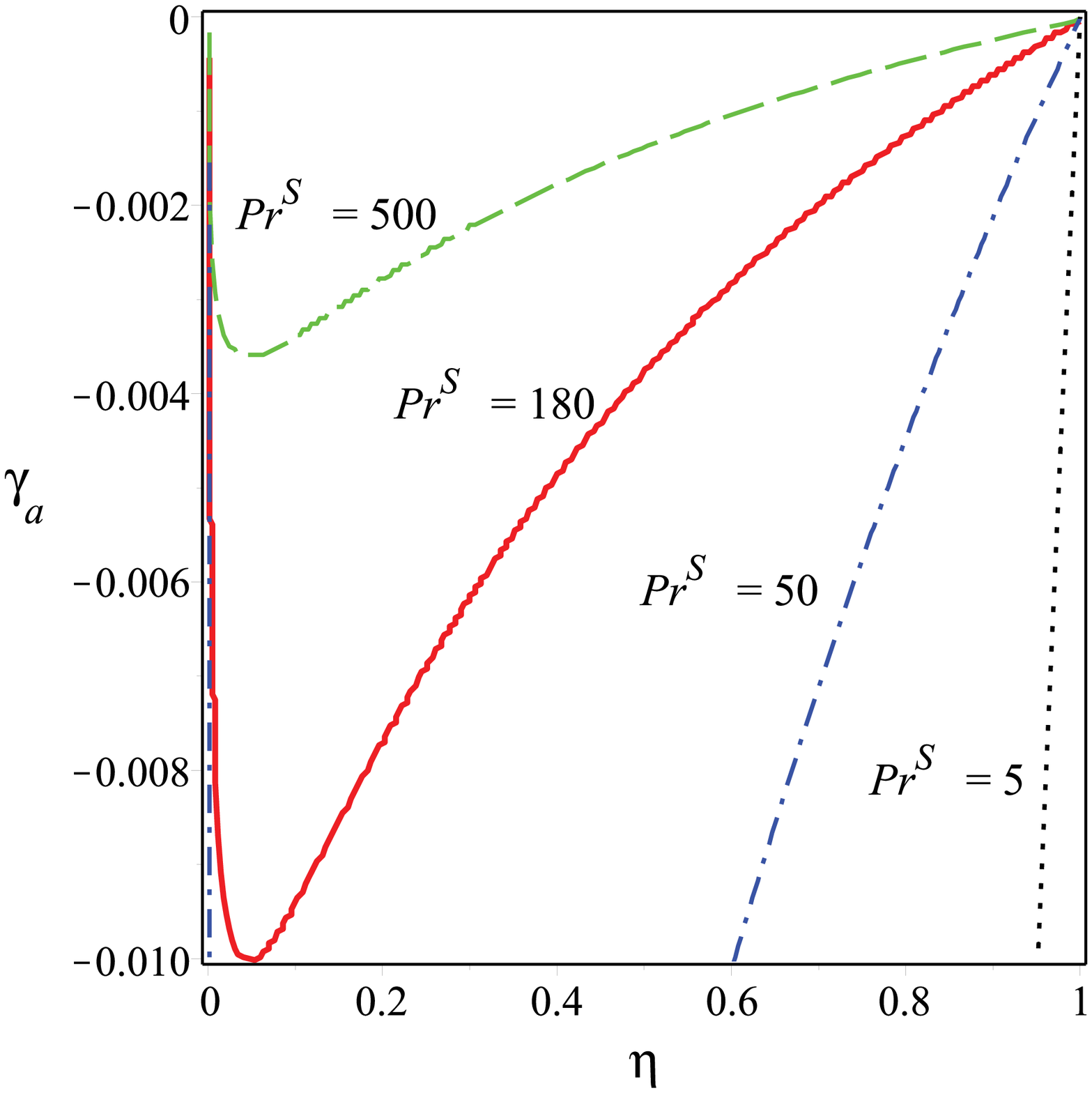}
    \includegraphics[angle=0, width=0.41\textwidth]{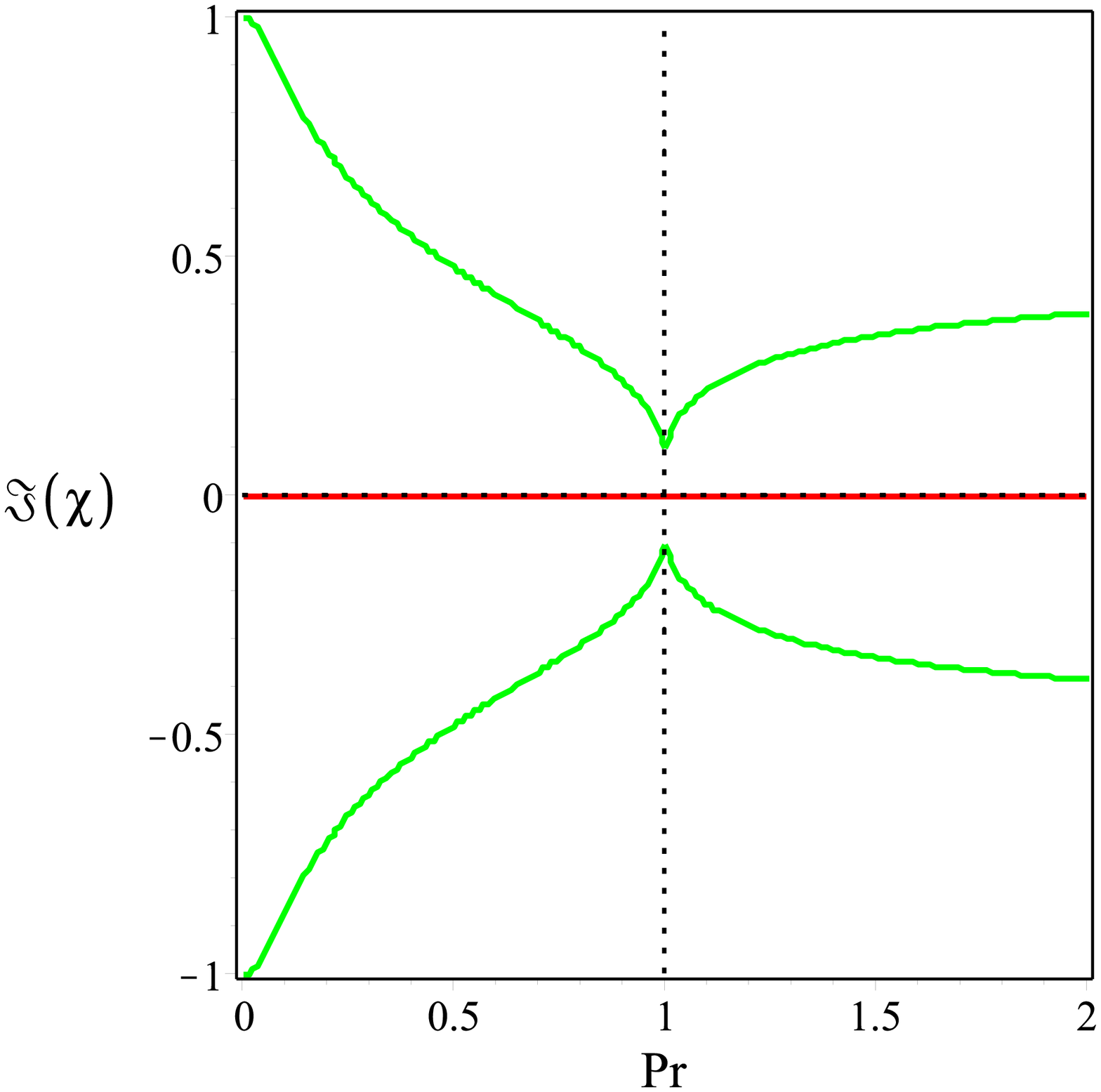}
    \includegraphics[angle=0, width=0.41\textwidth]{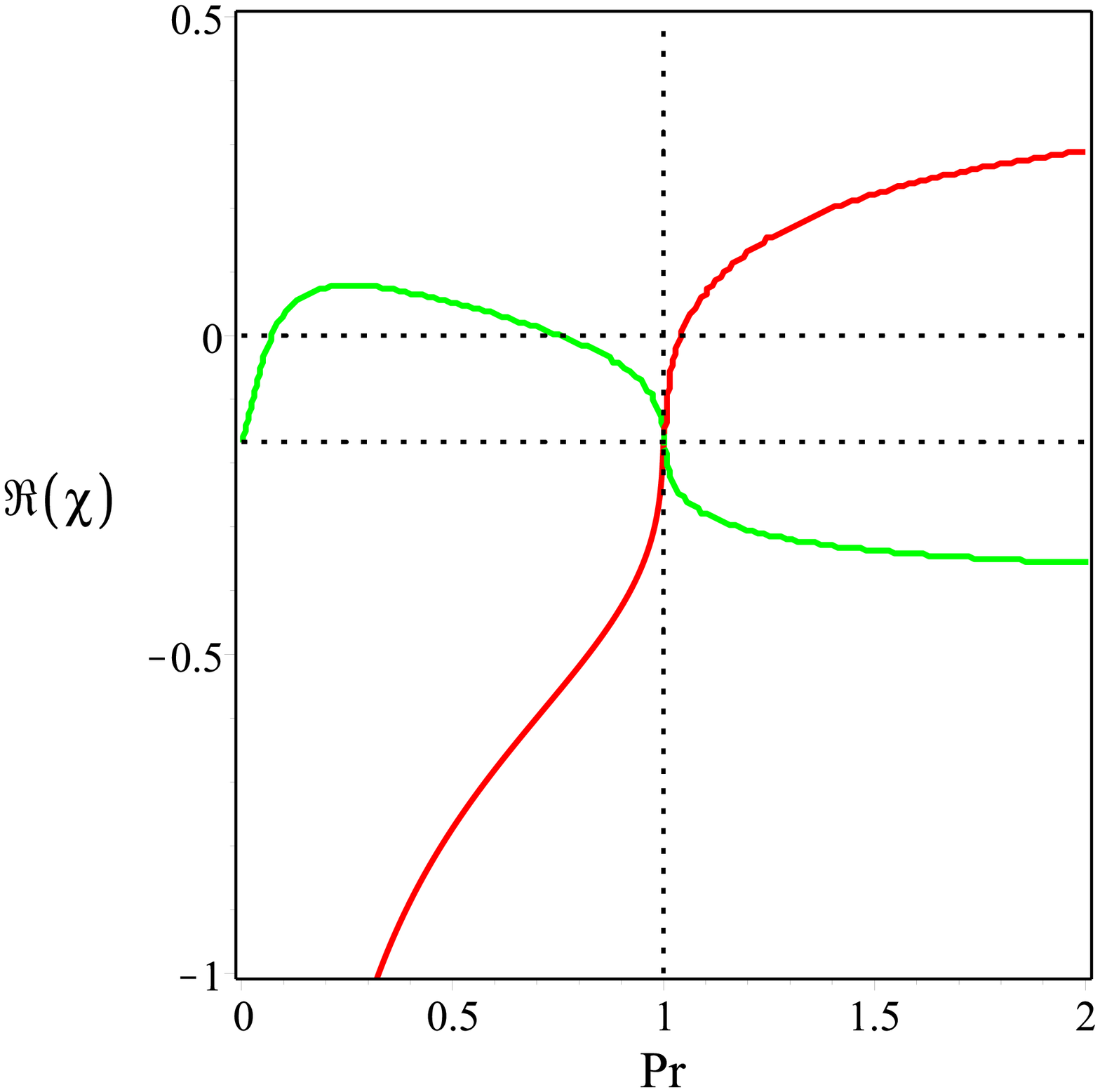}
    \end{center}
    \caption{Keplerian case with $\mu=\eta^{3/2}$ and $\eta=0.99$: (Upper left) Variation of the threshold of the stationary instability and Hopf bifurcation with $\rm Pr$   at $\gamma_a<0$ (the inward temperature gradient). The asymptotic values of $\rm Pr$ marked by dot-dashed lines are, respectively: ${\rm Pr}^{H}\approx0.0523$, ${\rm Pr}^{H}\approx0.98$, ${\rm Pr}^{S}\approx 1.01$ and ${\rm Pr}^{S}\approx 10.055$. (Upper right) Lines  ${\rm Pr}^{S}=const.$ in the $(\eta,\gamma_a)$--plane. (Lower row) Variation of the frequencies $({\rm Im}\chi)$ and the growth rates $({\rm Re}\chi)$ with $\rm Pr$ at ${\rm Ta}=6$ and $\gamma_a=-0.01$ (the inward temperature gradient). The green lines correspond to oscillatory modes, and the red lines to the stationary modes. }
    \label{fig3}
    \end{figure}

The stationary modes can exist if the radicand in \rf{stmodes} is positive, this yields a condition on ${\rm Pr}$ to be satisfied.  In the case of the Rayleigh unstable flow, i.e. when $1+{\rm Ro} < 0$, the stationary modes exist for any value of ${\rm Pr}$  when $\gamma_a < 0$ and for $0\le {\rm Pr} < {\rm Pr}^{S} $ when $\gamma_a > 0$, where the critical value of the Prandtl number is
\be{statmodcond}
{\rm Pr}^{S}=\frac{2(1+{\rm Ro})(1-\gamma_a\hat{\Theta})}{\gamma_a\hat{\Theta}{\rm Rt}}.
\ee

The values of ${\rm Pr}^{S}$ calculated in terms of the parameters \rf{meanpar} for different rotating cases of the Couette-Taylor system are given in Table~1. In the case of the rotating inner cylinder shown in the left panel of Figure~\ref{fig1} the asymptotic value \rf{statmodcond} for the threshold \rf{CTSA} is ${\rm Pr}^{S}\approx 9948$.

               \begin{figure}
    \begin{center}
    \includegraphics[angle=0, width=0.41\textwidth]{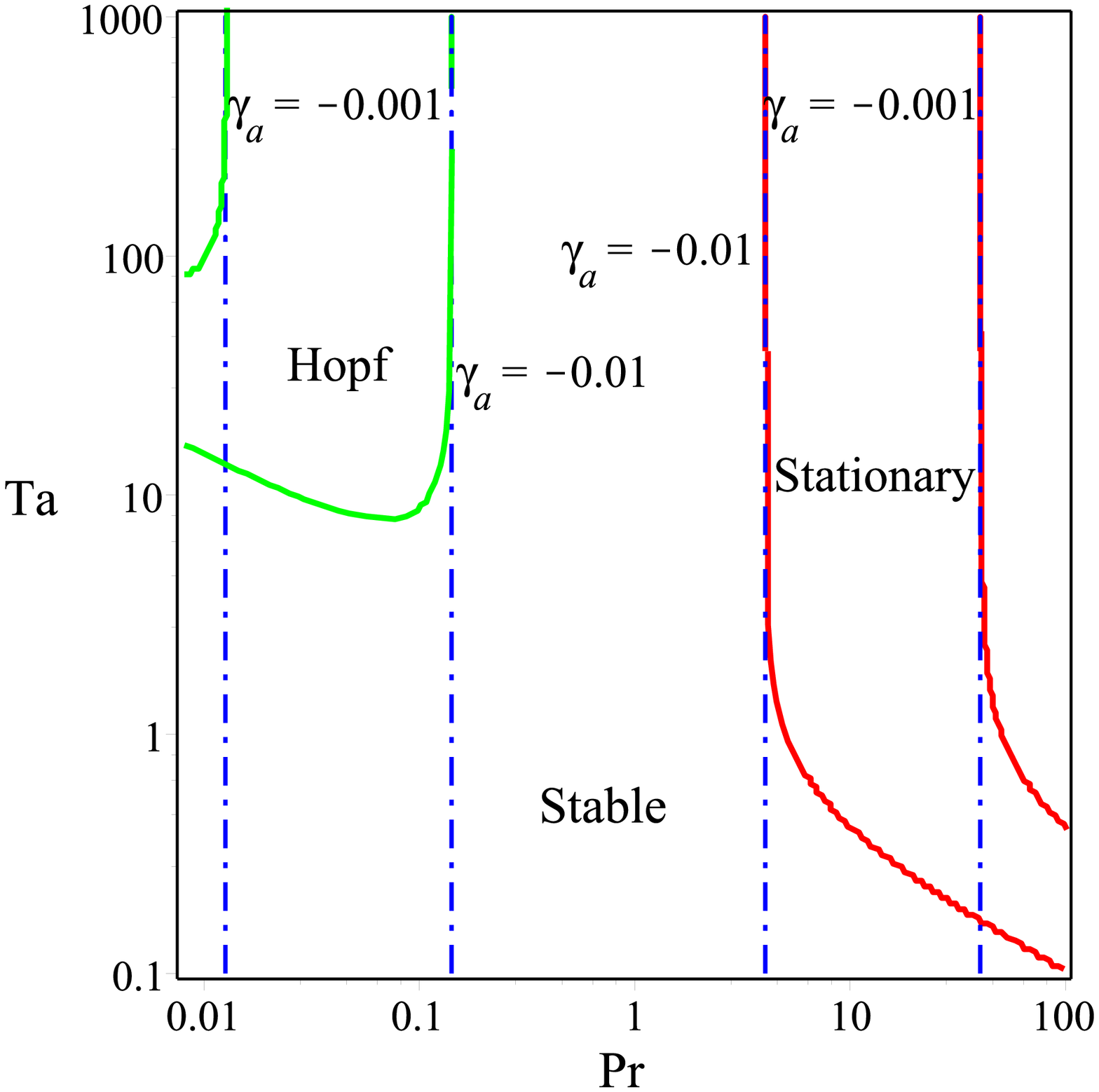}
    \includegraphics[angle=0, width=0.41\textwidth]{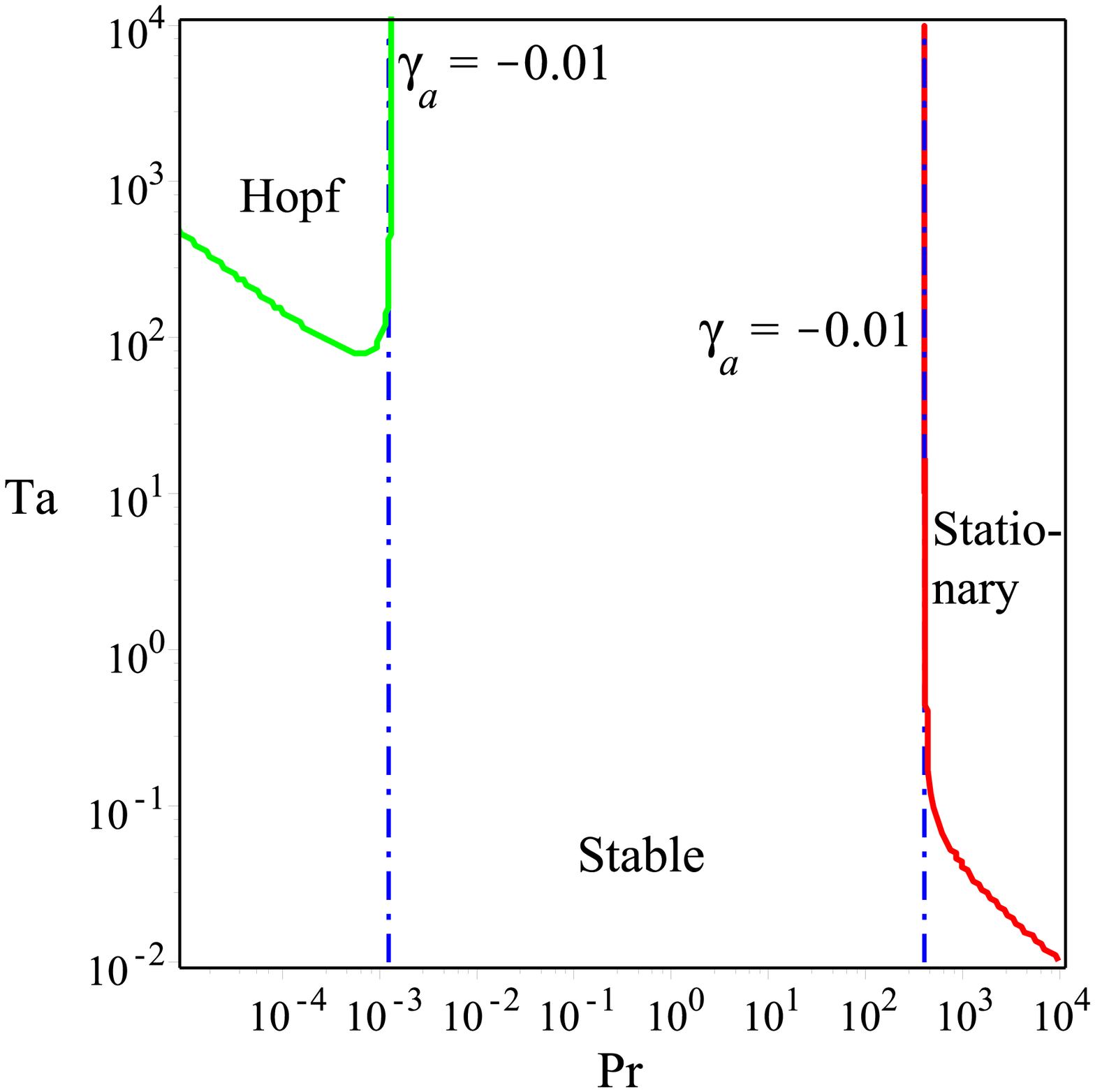}
    \includegraphics[angle=0, width=0.41\textwidth]{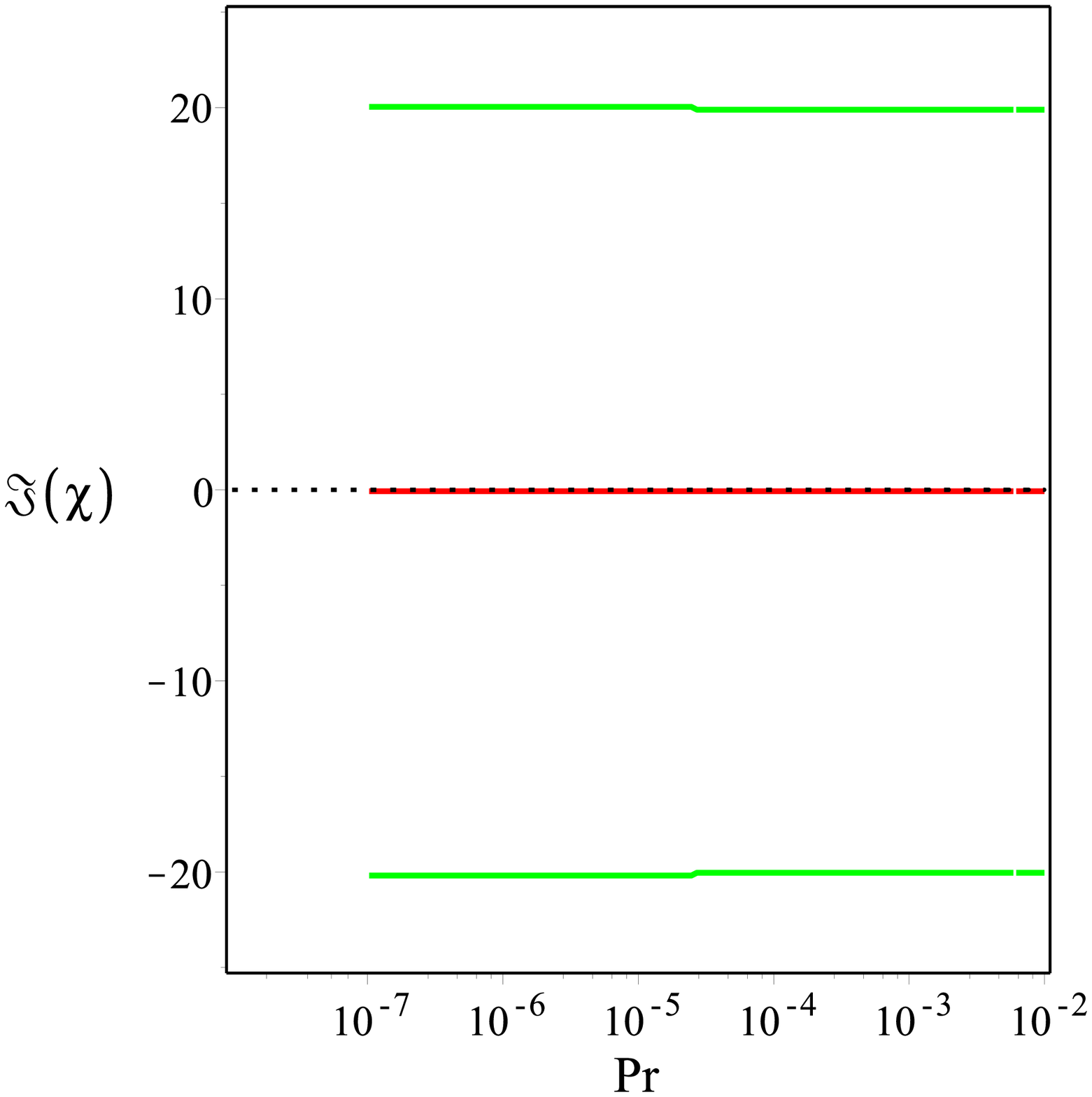}
    \includegraphics[angle=0, width=0.41\textwidth]{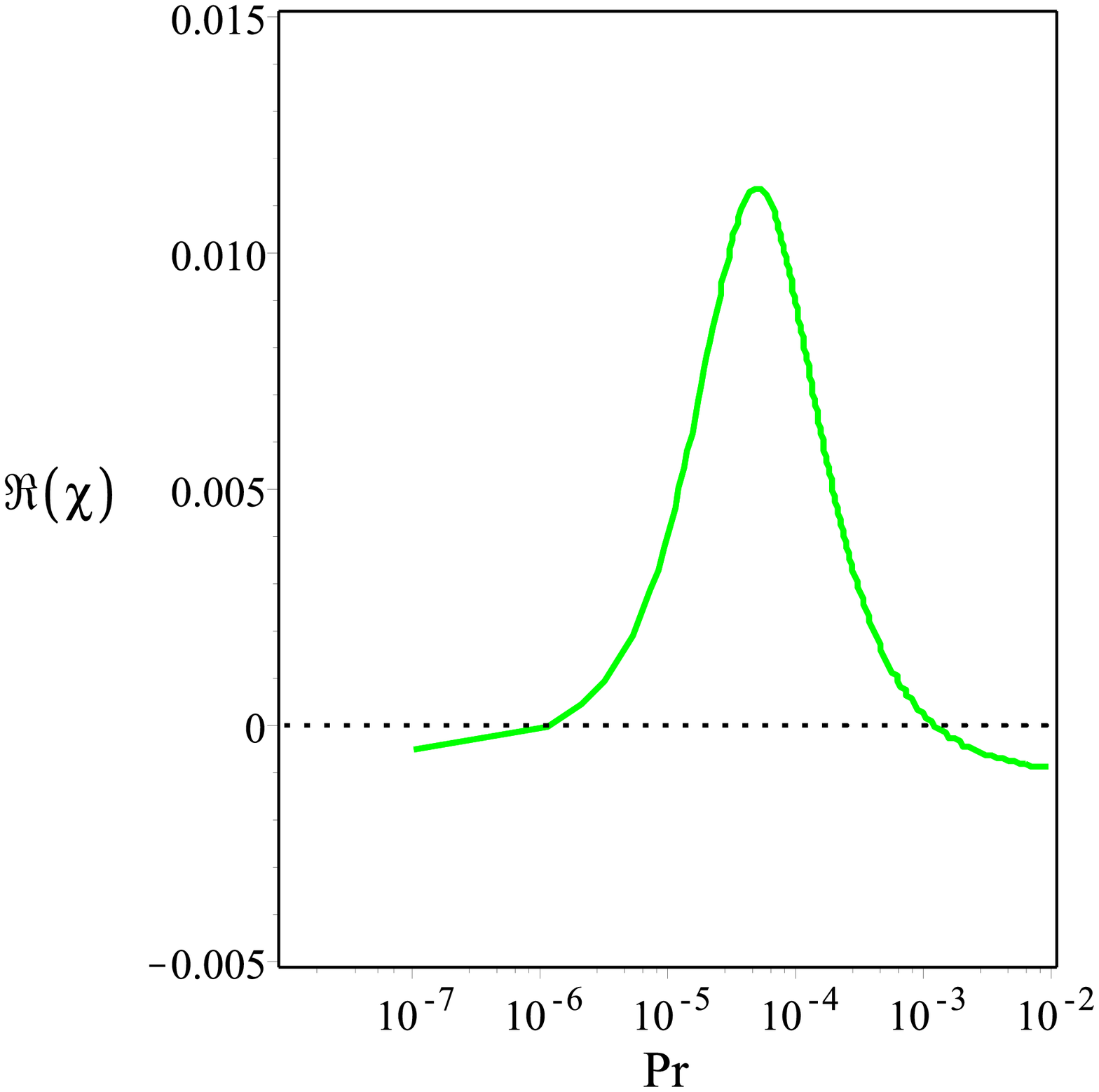}
    \end{center}
    \caption{(Upper left) Solid body rotation with $\mu=1$ and $\eta=0.99$. Thresholds of the stationary instability and Hopf bifurcation with the asymptotic values of $\rm Pr$ marked by dot-dashed lines: ${\rm Pr}^{H}\approx 0.0126$, ${\rm Pr}^{H}\approx0.1412$, ${\rm Pr}^{S}\approx 4.0402$, and ${\rm Pr}^{S}\approx 40.221$.
     (Upper right) Outer cylinder rotation with $\mu=\infty$ and $\eta=0.99$: Thresholds of the stationary instability and Hopf bifurcation with the asymptotic values of the Prandtl number: ${\rm Pr}^{H}\approx 0.0012$ and ${\rm Pr}^{S}\approx 404.02$. (Lower row) Variation of the (left) frequencies and (right) growth rates with $\rm Pr$ for the outer cylinder rotation ($\mu=\infty$, $\eta=0.99$) at given ${\rm Ta}=1000$ and $\gamma_a=-0.01$. The green lines correspond to the oscillatory modes and the red one corresponds to a (damped) stationary mode.}
    \label{fig4}
    \end{figure}

In the case of the Rayleigh stable flows $(1+{\rm Ro} >  0)$, the stationary modes are expected only for inward heating ($\gamma_a < 0$) at ${\rm Pr} > {\rm Pr}^{S}$. Figure~\ref{fig3} shows the variation of the threshold with ${\rm Pr}$ for the Keplerian rotation in the small gap case $\eta=0.99$; the corresponding stability diagrams for the cases of the solid body rotation  and the outer cylinder rotation are shown in Figure~\ref{fig4}.

Note that for the Rayleigh unstable flows the stationary modes exist even in the limit of
the vanishing  Prandtl number (${\rm Pr}\rightarrow 0$), i.e. for very well thermally conducting liquids with low kinematic viscosity. In this limit the threshold \rf{stmodes} reduces to the expression
$$ {\rm Ta}_c^{S} = \frac{1}{\sqrt{-4(1+{\rm Ro})(1-\gamma_a \hat{\Theta})}}$$  that does not depend on the sign of $\gamma_a$.

For the Rayleigh-stable flows  $(1+{\rm Ro}>0)$ the threshold of the stationary modes \rf{stmodes} decreases towards zero at $\gamma_a<0$ in the limit of large values of the Prandtl number (${\rm Pr}\rightarrow \infty$), i.e. for highly viscous and poorly heat-conducting fluids, see Figure~\ref{fig3} and \ref{fig4}.

\subsection{Hopf bifurcation}

In terms of the parameters \rf{meanpar} and \rf{iso} the threshold for the onset of oscillatory instability \rf{Hopf} becomes
\ba{TaSAOAGamma}
r_{H}=\frac{{\rm Ta}_c^{H}}{{\rm Ta}_0} &=&  \frac{{\rm Pr}+1}{\rm Pr}  {\left[ 1-\frac{\gamma_a}{2}+\frac{\gamma_a}{2 {\rm ln} \eta}\frac{{\rm Pr}+1}{{\rm Pr}}\left({\rm Ta}_0\right)^2\right]}^{-1/2}
\ea
and the Hopf frequency of the oscillatory modes \rf{imagroot} acquires the form
\be{FrqGamma}
\frac{\omega_c}{\Omega} = \frac{\beta}{{\rm Pr} + 1}\frac{1}{{\rm Ta}_0}\sqrt{-1+\frac{\gamma_a}{2}
 - \frac{\gamma_a{\rm Pr}({\rm Pr}+1)}{2 {\rm ln} \eta} \left({\rm Ta}_0\right)^2}.
\ee

                    \begin{figure}
    \begin{center}
    \includegraphics[angle=0, width=0.32\textwidth]{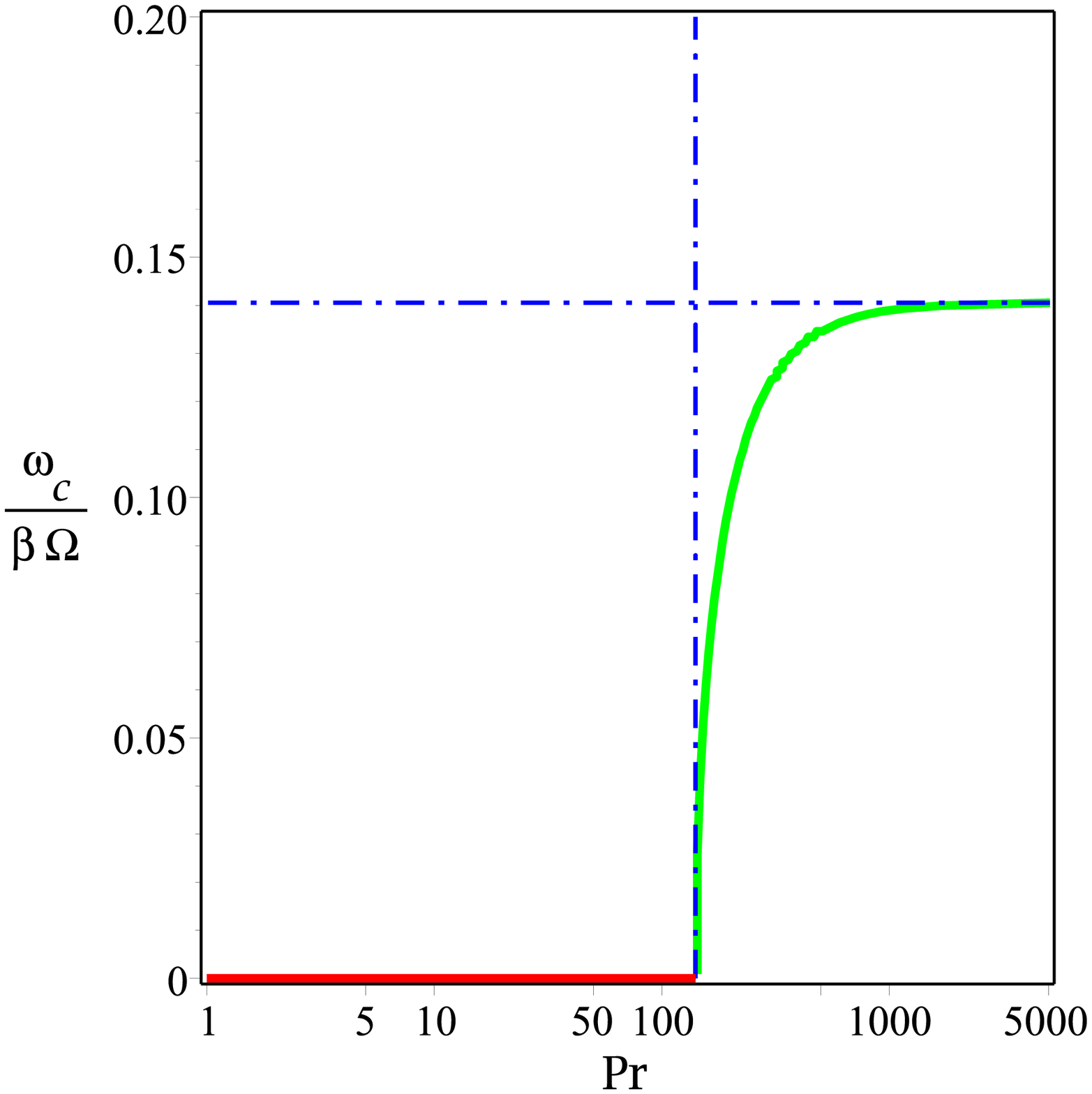}
    \includegraphics[angle=0, width=0.32\textwidth]{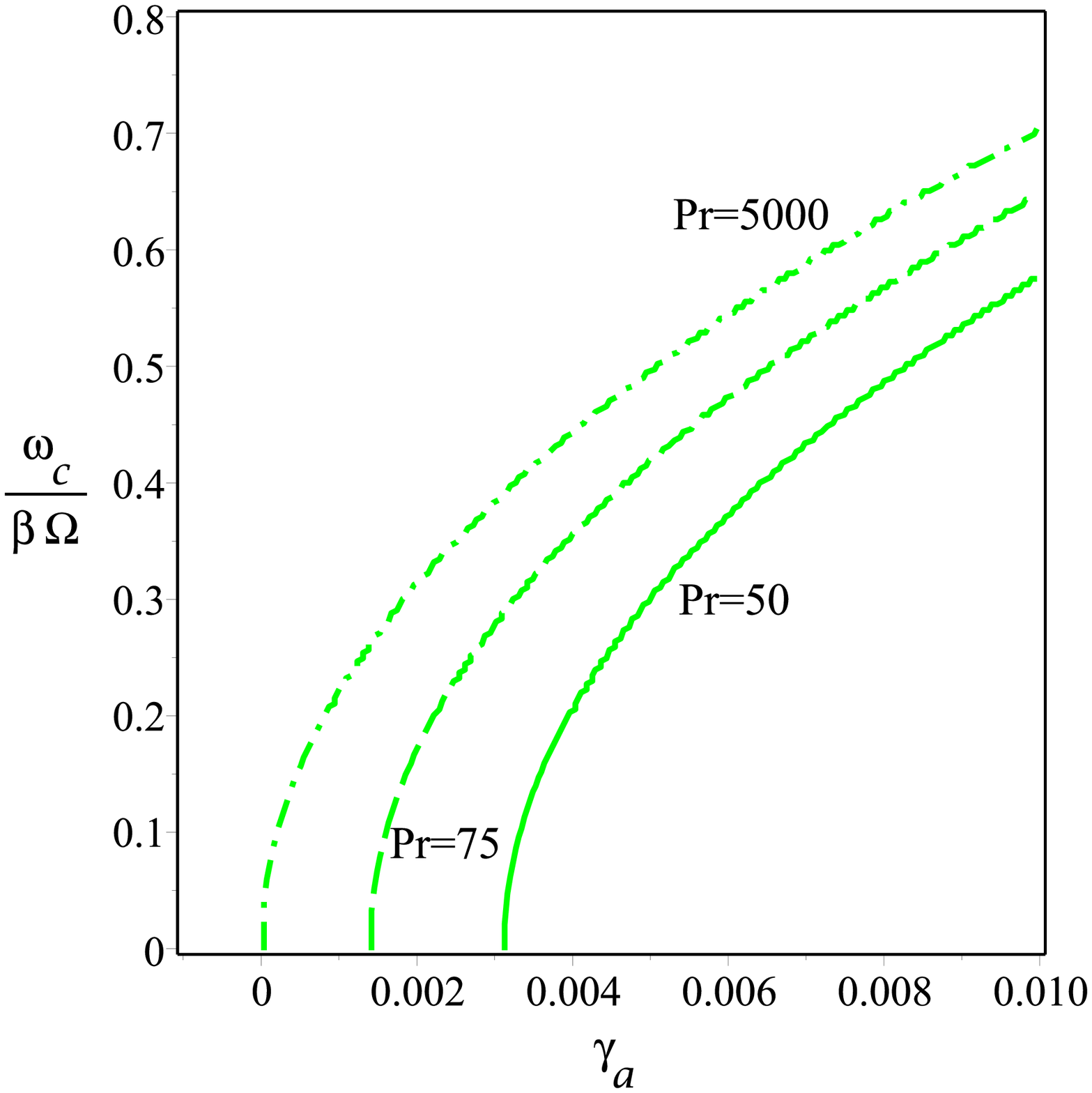}
    \includegraphics[angle=0, width=0.32\textwidth]{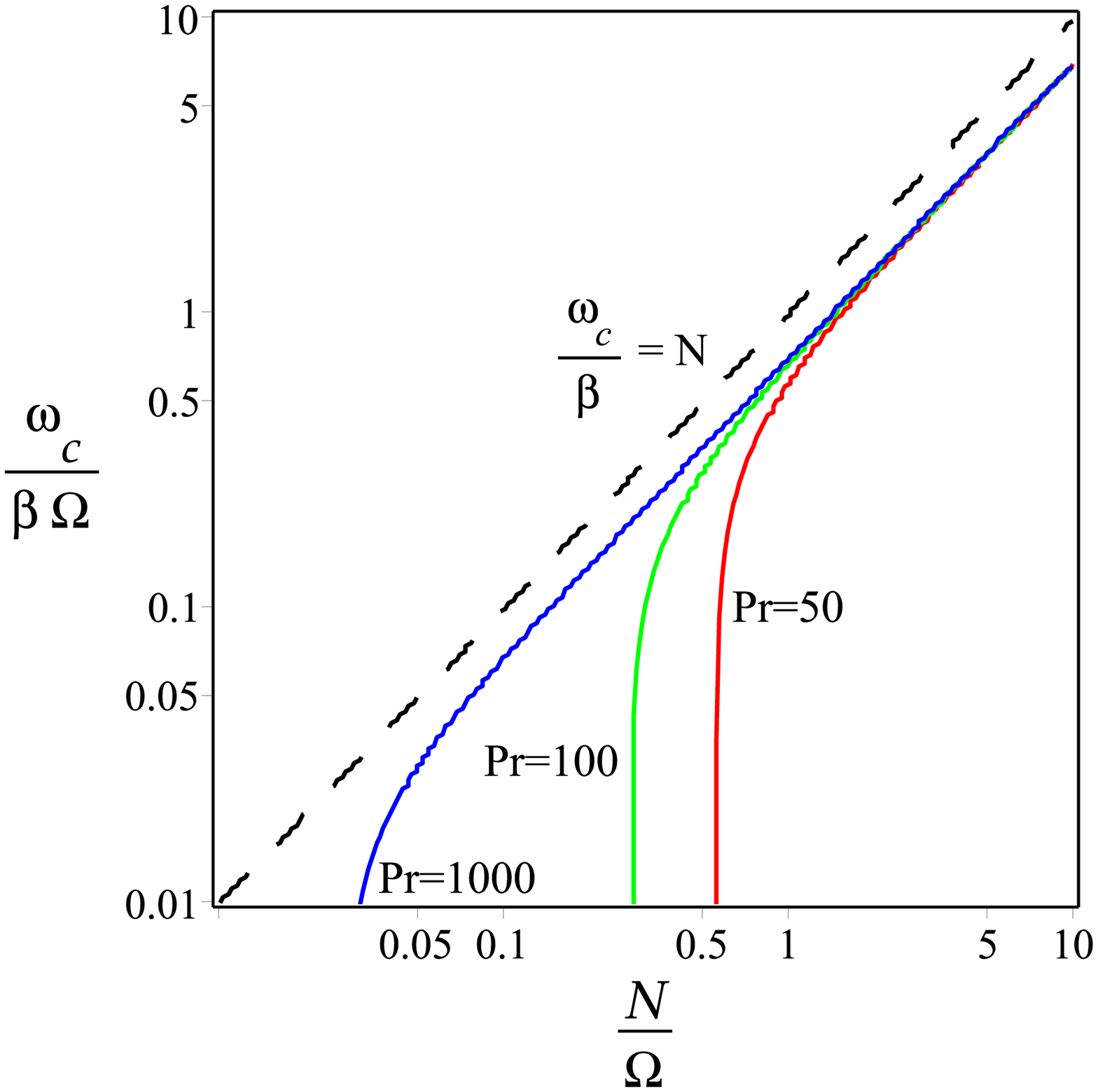}
    \end{center}
    \caption{Rotating inner cylinder with $\mu=0$ and $\eta=0.99$. The critical frequency $\omega_c$ at the onset of (red) stationary instability and (green) Hopf bifurcation in the units of $\beta \Omega$ (left) versus $\rm Pr$ at $\gamma_a=0.0004$ and (center) versus $\gamma_a$ according to Eq.~\rf{FrqGamma}. The limiting frequency \rf{omin} at $\gamma_a=0.0004$ is $\frac{\omega_{\infty}}{\beta\Omega}\approx0.1411$. (Right) The critical Hopf frequency $\omega_c$ as a function of the Brunt-V\"ais\"al\"a frequency $\rm N$  according to Eq.~\rf{frq} and Eq.~\rf{bvfln}. The critical frequency vanishes at ${\rm N}_c/\Omega\approx0.5572$ for $\rm Pr=50$, at ${\rm N}_c/\Omega\approx0.28$ for $\rm Pr=100$, and at ${\rm N}_c/\Omega\approx0.0281$ for $\rm Pr=1000$ calculated by means of Eqs.~\rf{BVCrit} and \rf{meanpar}. }
    \label{fig2}
    \end{figure}

The left and the central panels of Figure~\ref{fig1} illustrate the variation of the threshold of the Hopf bifurcation with ${\rm  Pr}$ and $\gamma_a$ for the circular Couette flow with the inner cylinder rotating alone. The variation of the Hopf frequency with ${\rm  Pr}$ and $\gamma_a$ in this rotating case is shown in Figure~\ref{fig2}.  Similarly to the stationary instability, both the threshold \rf{TaSAOAGamma} and the frequency \rf{FrqGamma} of the oscillatory modes depend explicitly on the Prandtl number ${\rm Pr}$ and on the temperature drop parameter $\widehat S$ in agreement with the numerical calculations of \citep{MYM2015}.

We see that at $\gamma_a > 0$ both $r_{H} > 1$ and $r_{S} > 1$, i.e. the outward heating is stabilizing for both stationary and oscillatory modes with respect to the isothermal flow. The growth rates shown in the right panel of Figure~\ref{fig1} demonstrate that the complex roots do not exist at $\gamma_a \le 0$; they appear at $\gamma_a >0$ due to splitting of a double real eigenvalue. When this eigenvalue is negative the complex roots are stable. They acquire a positive growth rate only after a double real eigenvalue passes through zero with the change of parameters, which happens exactly at a codimension-2 point, cf. \cite{LT2001}.

Indeed, since this flow is Rayleigh-unstable in the absence of diffusion, the branches of the stationary and oscillatory modes have a common codimension-2 point with the coordinates \rf{codim2a} and \rf{codim2ab} that in the parametric representation \rf{meanpar} and \rf{iso} can be written as follows:
\ba{CD2pt}
{\rm Pr}^{*} &=&-\frac{1}{2}+\frac{1}{2}\sqrt{1-4\frac{2-\gamma_a}{\gamma_a}\frac{\ln{\eta}}{ ({\rm Ta}_0)^2}}, \\
{\rm Ta}^{*} &=& {\rm Ta}_0\left[1-\frac{\gamma_a}{2}+ \gamma_a{\rm Pr}^*\frac{\left( {\rm Ta}_0 \right)^2}{{\rm ln} \eta}\right]^{-1/2}.\label{CD2pta}
\ea
At the codimension-2 point one of the roots of the dispersion relation is double zero and the other one is  simple real and negative, corresponding to a damped mode, as the growth rates demonstrate in the right panel of Figure~\ref{fig1}.

The threshold for the excitation of oscillatory modes and the corresponding critical frequency of the oscillatory modes for both the Rayleigh-stable and the Rayleigh-unstable rotation regimes in the Couette-Taylor system are summarized in Table 2. For the Rayleigh unstable flows with the outward heating the oscillatory instability actually occurs at ${\rm Pr}>{\rm Pr}^*$, where ${\rm Pr}^*$ is given by Eq.~\rf{codim2a} or Eq.~\rf{CD2pt}, see Figure~\ref{fig1}.

Figure~\ref{fig3} and \ref{fig4} show the boundaries of stationary instability and Hopf bifurcation in the $({\rm Pr},{\rm Ta})$ - plane for the Rayleigh-stable flows such as Keplerian rotation, solid-body rotation and the case of the outer cylinder rotating. We see also that the frequency of the oscillatory instability is increasing while approaching the outer cylinder rotating case whereas the growth rate of the oscillatory instability at low $\rm Pr$ is decreasing.

All these Rayleigh stable flows ($1+{\rm Ro}>0$ or $\mu \ge {\eta}^2$) with the inward heating $(\gamma_a<0)$ share the same property. Namely, the oscillatory modes are excited when thermal diffusion dominates over the fluid viscosity
\be{Hopfcond}
0<{\rm Pr} < {\rm Pr}^{H}=\frac{\gamma_a\hat{\Theta}{\rm Rt}}{\gamma_a\hat{\Theta}{\rm Rt}-4(1+{\rm Ro})(1-\gamma_a\hat{\Theta})}<1.
\ee
This result follows from the observation that the Hopf bifurcation from the base flow can occur if the radicand in \rf{Hopf} is positive. In contrast, the stationary modes are excited in the Rayleigh-stable flow by the inward heating when ${\rm Pr}>{\rm Pr}^S>1$, where ${\rm Pr}^S$ is given in \rf{statmodcond} and Table~1.

Therefore, for the Rayleigh stable flows the inward radial temperature gradient destabilizes the flow depending on the value of ${\rm Pr}$. The oscillatory instability occurs at $0<{\rm Pr}<1$ while the stationary instability takes place at ${\rm Pr}>1$.  At ${\rm Pr}=1$ the flows remain stable, becoming unstable only when ${\rm Pr}\ne 1$ which is a characteristic feature of the double-diffusive instabilities \citep{AG1978,KV2010,K2013}.

As Figure~\ref{fig3} and \ref{fig4} demonstrate, when ${\rm Pr}\rightarrow 0$, the threshold of the oscillatory modes \rf{Hopf} occurring in the Rayleigh stable flows becomes infinite
$({\rm Ta}_c^H \rightarrow \infty)$ while their frequency $\bar\omega_c \rightarrow 2\sqrt{(1+{\rm Ro})(1-\gamma_a\hat\Theta)}$ depends weakly on the heating parameter $\gamma_a$.

The opposite limit, when ${\rm Pr}\rightarrow \infty$, is applicable to the threshold of the Hopf bifurcation in the Rayleigh unstable flows with the outward heating $(\gamma_a>0)$, yielding ${\rm Ta}_c^H \rightarrow [\gamma_a\hat{\Theta}{\rm Rt}-4(1-\gamma_a\hat\Theta)(1+{\rm Ro})]^{-1/2}$. The corresponding Hopf frequency becomes independent on the rotation rate: $\bar{\omega}_{\infty}= \sqrt{-\gamma_a\hat{\Theta}{\rm Rt}}$ as is visible in the left panel of Figure~\ref{fig2}. The central panel of the same Figure shows that  the Hopf frequency  is bounded $(0 < \bar{\omega}_c< \bar{\omega}_{\infty})$ at any fixed
${\rm Pr}^*<{\rm Pr}< \infty$.

The frequency $\bar{\omega}_{\infty}$ is fixed by the  Brunt-V\"ais\"al\"a frequency $\rm N$ in the case of stable stratification of the fluid density ($\gamma_a>0$, ${\rm Rt}<0$), as is evident in the right panel of Figure~\ref{fig2}.
Indeed, for a fixed value of ${\rm Pr}$, the Hopf frequency \rf{imagroot} can be expressed via the   Brunt-V\"ais\"al\"a frequency as
\be{frq}
\bar{\omega}_c=\sqrt{\frac{\rm N^2}{2\Omega^2}\frac{\rm Pr}{{\rm Pr}+1}+\frac{4(1+{\rm Ro})(1-\gamma_a\hat{\Theta})}{({\rm Pr}+1)^2}}.
\ee
Hence, in the limit of ${\rm Pr} \rightarrow \infty$ we have
\be{omin}
\bar \omega_{\infty}=\frac{1}{\sqrt{2}}\frac{\rm N}{\Omega}=\sqrt{-\frac{\gamma_a}{2{\rm ln}{\eta}}},
\ee
which, in particular, yields another expression for the Brunt-V\"ais\"al\"a frequency in the case of the Rayleigh unstable flows with the outward heating $(\gamma_a > 0)$:
\be{bvfln}
{\rm N}^2=-\frac{\gamma_a}{\ln \eta}\Omega^2.
\ee

The right panel of Figure~\ref{fig2} shows that all the critical frequencies tend asymptotically to the line \rf{omin} as $\rm N/\Omega \rightarrow \infty$. With the decrease in the ratio $\rm N/\Omega$ the critical frequency vanishes at the characteristic value of the  Brunt-V\"ais\"al\"a frequency $\rm N_c$ given by
\be{BVCrit}
\frac{{\rm N}_c^2}{\Omega^2} =-\frac{8(1+{\rm Ro})(1-\gamma_a \hat{\Theta})}{{\rm Pr}({\rm Pr}+1)}.
\ee
The characteristic Brunt-V\"ais\"al\"a frequency $\rm N_c$  determines the stratification that is required to excite the inertial waves by the rotation. Since $\rm N_c$ decreases with the increase in ${\rm Pr}$, Figure~\ref{fig2}, we conclude that the inertial waves are easier to excite in the Rayleigh-unstable flow by the outward radial temperature gradient in the case of highly viscous fluids than in the case of the weakly viscous fluids.

\section{Discussion}

The short wavelength approximation has been used to investigate the stability of the Taylor-Couette flow with a radial buoyancy force induced by the coupling between radial temperature gradient and the centrifugal force in the Boussinesq approximation. A characteristic polynomial equation has been analyzed and marginal stability branches of stationary and oscillatory modes have been found analytically. The present method allows one to find the explicit dependence of the marginal state with the main control parameters: ${\rm Pr}$, ${\rm Ro}$, $\gamma_a$, and ${\rm Rt}$.

The reader should be aware that for the validity of the Boussinesq approximation, the thermal expansion parameter must be very small i.e. $| \gamma_a| \ll 1$ so that it is possible to neglect $\gamma_{a}$ before $1$  in all expressions containing $1-\gamma_{a}/2$. In fact, in our computations, the largest value of $| \gamma_a|$ was $| \gamma_a| = 0.01$ for fluids with the largest expansion coefficient found in literature to which a maximum of temperature difference applied should not exceed $\Delta T_{max} = 5{^\circ}C$.

From the above results computed at the geometric mean radius,  for each value of rotating case of the circular Couette flow (i.e. for a given value of $\mu$), it is possible to determine the critical modes and their threshold ${\rm Ta}_c$ as a function of different flow parameters $\eta,  \gamma_a$, and ${\rm Pr}$ together with the critical frequency for oscillatory modes. We have determined the critical parameters for the particular case of specific interest: the inner cylinder sole rotation ($\mu = 0$), the outer cylinder sole rotation ($\mu \rightarrow \infty$), the solid body rotation ($\mu = 1$) and the Keplerian rotation ($\mu = \eta^{3/2}$).

For the inner cylinder sole rotation, we have retrieved and extended the results of linear stability analysis \citep{MYM2015}. For example, we have  proved easily that  the slope of the branch of stationary modes at the origin ($\gamma_a=0$) is increasing with both  ${\rm Pr}$ and $\eta$. In fact, one gets
\be{der}
\frac{d{\rm Ta}}{d\gamma_a}(\gamma_a=0)= \frac{{\rm Ta}_0}{4} \left(1-\frac{2{\rm Pr}({\rm Ta}_0)^2 }{{\rm ln}\,\eta}\right).
\ee

To our best knowledge, the Rayleigh stable flows (except the solid body rotation) with a radial temperature gradient have never been treated analytically in such a detail so far and the results presented in this work offer new predictions of convective instabilities. In the case of the outer cylinder sole rotation, in the case of the solid body rotation, and in the Keplerian regime the inward heating is destabilizing and leads to the oscillatory instability via Hopf bifurcation at $0<{\rm Pr}<{\rm Pr}^{H}<1$ and to stationary instability for any value ${\rm Pr} > {\rm Pr}^{S}>1$  where the thresholds depend on $\gamma_a$ and $\eta$ (Table 1). The values of ${\rm Pr}$ for which the oscillatory instability is observed are comparable to that of the astrophysical flows for which ${\rm Pr} \approx 0.02$. At the Rayleigh line the flow is unstable for inward heating. The present model is based on the short wavelength approximation in the axial direction, so, strictly speaking, it cannot be applied as such to the case of the solid body rotation where it is known \citep{Auer1995} that the critical modes are columnar vortices ($n \neq 0, k=0$).

In the Appendix we have compared the present results with those derived by Economides and Moir \citep{EM1980} for the Taylor-Couette flow with a centripetal acceleration and a radial temperature gradient.  \cite{Lopez2013} have investigated the stability of the rotating cylindrical annulus with a negative radial temperature gradient and in the presence of the gravity $g$. While these authors suggested that quasi-Keplerian flows may be stable for weak stratification in the radial direction, our results show that laminar quasi-Keplerian flows may be destabilized by the centrifugal buoyancy, leading to oscillatory modes for values of ${\rm Pr}$ relevant to accretion-disc problem, although we have considered a model that has solid radial boundaries.



\section{Conclusion}

The short wavelength approximation  method has been applied to the linear stability of the circular Couette flow with a centrifugal buoyancy induced by the  radial temperature gradient in the absence of the natural gravity. The marginal states (stationary and oscillatory modes) have been determined analytically and the effects of the different parameters of the problem on the flow stability have been analyzed in detail. The centrifugal buoyancy enhances the instability of Rayleigh unstable flows in the inward heating and in the outward heating it induces oscillatory modes the frequency of which can be compared with the Brunt-V\"ais\"ala frequency in the limit of large ${\rm Pr }$ values. In the inward heating it also induces instability in Rayleigh stable flows in the form of stationary modes for large values of ${\rm Pr }$ or oscillatory modes for small values of ${\rm Pr }$. The present study may serve as a theoretical guideline to linear stability analysis and to direct numerical simulations (DNS) of the flow.

\section*{{Acknowledgments}}

O.N.K. has benefited from the CNRS grant for visiting senior scholars and acknowledges financial support from the ERC Advanced Grant ``Instabilities and nonlocal multiscale
modelling of materials'' FP7-PEOPLE-IDEAS-ERC-2013-AdG (2014-2019). I.M. is thankful to A. Meyer and Y. Harunori and O.N.K. is thankful to L. Tuckerman for fruitful exchanges on this work. I.M. acknowledges the support from the French National Research Agency (ANR) through the program Investissements d'Avenir (No. ANR-10 LABX-09-01), LABEX EMC$^3$ (project TUVECO) and from the CPER Normandie.

\appendix
\section{Appendix. Link to \cite{EM1980} }

In the Appendix we would like to compare our short-wavelength equations with the matrix \rf{matrix}
to the results by Economides and Moir \citep{EM1980} derived in the narrow gap approximation.

Let $d=R_2-R_1 \ll R_1$ is the size of the gap between the cylinders in the Couette cell.
Then, we can write
\be{ecom7}
x=(r-R_1)/d,\quad \mu=\Omega_2/\Omega_1 ,\quad \Omega(x)=1-(1-\mu)x, \quad x\in [0,1].
\ee
Denoting $D=d/dx$ so that $d/dr=d^{-1}d/dx$ we define the operators
\ba{ecom9}
L&=&D^2-(k_z d)^2-\sigma-ik\sqrt{T} \Omega(x),\nn\\
M&=&D^2-(k_z d)^2-\sigma {\rm Pr} -ik{\rm Pr}\sqrt{T}\Omega(x),
\ea
where ${\rm Pr}=\nu/\kappa$ is the Prandtl number and
\be{ecom8}
\sigma=\frac{\omega d^2}{\nu},\quad k=m\sqrt{-\frac{\Omega_1}{4A}},\quad T=-\frac{4A\Omega_1d^4}{\nu^2},\quad \bar\beta=\frac{T_2-T_1}{R_2-R_1}.
\ee
The temporal eigenvalue is denoted by $\omega$ and the azimuthal wavenumber by $m$.
The parameter $A$ is a coefficient in the expression for the background circular Couette flow
\be{ecom5}
r\Omega(r)=A r + \frac{B}{r}
\ee
and can be expressed via the Rossby number $\rm Ro$ as \cite{KSF2014}:
\be{ecom6}
A=\Omega(1+{\rm Ro}).
\ee

Then, the system of the narrow gap equations derived in \cite{EM1980} is
\ba{ecom2}
L(D^2-(k_z d)^2)u'&=&-(k_z d)^2T\Omega(x)v'+(k_z d)^2R\theta',\nn\\
Lv'&=&u',\nn\\
M\theta'&=&u',
\ea
where (with $g$ denoting constant gravity acceleration and $\alpha$ the coefficient of thermal expansion)
\be{ecom3}
R=\frac{g\alpha \bar\beta d^4}{\nu \kappa}, \quad u'=\frac{2Ad \delta}{\nu \Omega_1} u,\quad v'=\frac{v}{R_1 \Omega_1},\quad
\theta'=\frac{2A \kappa \theta}{\bar\beta \nu \Omega_1 R_1},\quad \delta=\frac{d}{R_1}.
\ee

For any $w \sim \exp(i k_r r)$, we have
\ba{ecom10}
Lw&=&- \frac{d^2}{\nu}\left[\nu(k_r^2+k_z^2)+\omega+im\Omega_1\Omega(x)\right]w\nn\\
&=&- \frac{d^2}{\nu}\left[\nu|{\bf k}|^2+\omega+im\Omega_1\Omega(x)\right]w\nn\\
&=&- \frac{d^2}{\nu}\left[\omega_{\nu}+\omega+im\Omega_1\Omega(x)\right]w.
\ea
Hence, if $u,v \sim \exp(i k_r r)$ the first of equations \rf{ecom2} becomes
\ba{ecom11}
- \frac{d^2}{\nu}\left[\omega_{\nu}+\omega+im\Omega_1\Omega(x)\right](-d^2|{\bf k}|^2)\frac{2Ad \delta}{\nu \Omega_1}u&=&-(k_z d)^2 (-4A \Omega_1 d^4)\frac{\Omega(x)}{\nu^2 R_1 \Omega_1}v\nn\\
&+&(k_z d)^2\frac{2A g \alpha d^4}{\nu^2 \Omega_1 R_1}\theta.
\ea
Simplifying the above expression, we find
\ba{ecom12}
\left[\omega_{\nu}+\omega+im\Omega_1\Omega(x)\right] u&=&\frac{k_z^2}{|{\bf k}|^2} 2 \Omega_1\Omega(x) \frac{d}{ \delta R_1 }v+\frac{k_z^2}{|{\bf k}|^2}\frac{  d}{ \delta R_1}g \alpha\theta
\ea
and finally
\ba{ecom13}
\left[\omega_{\nu}+\omega+im\Omega_1\Omega(x)\right] u&=&\frac{k_z^2}{|{\bf k}|^2} 2 \Omega_1\Omega(x) v+\frac{k_z^2}{|{\bf k}|^2}g \alpha\theta,
\ea
where $\omega_{\nu}=\nu|{\bf k}|^2$.

Analogously, the second of equations \rf{ecom2} becomes
\be{ecom14}
- \frac{d^2}{\nu R_1 \Omega_1}\left[\omega_{\nu}+\omega+im\Omega_1\Omega(x)\right]v=\frac{2Ad \delta}{\nu \Omega_1} u.
\ee
Simplifying it, we get
\be{ecom15}
\left[\omega_{\nu}+\omega+im\Omega_1\Omega(x)\right]v=-2A  u
\ee
and, finally
\be{ecom16}
\left[\omega_{\nu}+\omega+im\Omega_1\Omega(x)\right]v=-2\Omega(1+{\rm Ro})  u.
\ee

Now, for the third of equations \rf{ecom2} we find
\be{ecom17}
\left[-k_r^2 d^2-(k_z d)^2-\sigma {\rm Pr} -ik{\rm Pr}\sqrt{T}\Omega(x) \right]\frac{2A \kappa}{\bar\beta \nu \Omega_1 R_1 }\theta=\frac{2Ad \delta}{\nu \Omega_1} u,
\ee
and equivalently,
\be{ecom18}
\left[-\kappa k_r^2 -\kappa k_z^2 -\omega  -im\Omega_1\Omega(x) \right]\frac{d }{\bar\beta \nu R_1 }\theta=\frac{ \delta}{\nu } u,
\ee
so that finally
\be{ecom18}
\left[\omega_{\kappa}+\omega  +im\Omega_1\Omega(x) \right]\theta= -\bar\beta u,
\ee
where $\omega_{\kappa}=\kappa|{\bf k}|^2$.

We can denote $\Omega=\Omega_1\Omega(x)$ and write the three equations in the matrix form
\be{ecomatrix}
\left(
  \begin{array}{ccc}
    \omega_{\nu}+\omega+im\Omega & -\frac{k_z^2}{|{\bf k}|^2} 2 \Omega & -\frac{k_z^2}{|{\bf k}|^2}g \alpha \\
    2\Omega(1+{\rm Ro}) & \omega_{\nu}+\omega+im\Omega & 0 \\
    \bar\beta  & 0 & \omega_{\kappa}+\omega  +im\Omega \\
  \end{array}
\right)\left(
         \begin{array}{c}
           u \\
           v \\
           \theta \\
         \end{array}
       \right)=0.
\ee
In fact, $\bar\beta=\frac{T_2-T_1}{R_2-R_1}$ is a `derivative' of the temperature with respect to radius and in local approximation we can replace it with $\Theta'=d\theta_0/dr$. Then, we can denote $\beta=\frac{k_z}{|{\bf k}|}$ and write
\be{ecomatrix}
\left(
  \begin{array}{ccc}
    \omega_{\nu}+\omega+im\Omega & - 2 \Omega \beta^2 & - \alpha g \beta^2\\
    2\Omega(1+{\rm Ro}) & \omega_{\nu}+\omega+im\Omega & 0 \\
    \Theta'  & 0 & \omega_{\kappa}+\omega  +im\Omega \\
  \end{array}
\right)\left(
         \begin{array}{c}
           u \\
           v \\
           \theta \\
         \end{array}
       \right)=0.
\ee
For the stationary axisymmetric instability the Bilharz criterion \citep{B1944} applied to the characteristic polynomial of the matrix in the left hand side of  equation \rf{ecomatrix} yields:
\be{emsac}
4({\rm Ro}+1)+{\rm Pr}\frac{g}{\Omega^2}\frac{2{\rm Rt}\Theta\alpha}{r}+\frac{1}{{\rm Ta}^2}>0
\ee
With $g=\Omega^2 r$, and taking into account that the sign of $\alpha$ in \citep{EM1980} is opposite to the sign of $\alpha$ in \rf{nsht}, we reduce the above equation to the form
\be{emsacf}
4({\rm Ro}+1)-2{\rm Pr}{\rm Rt}\hat\Theta\gamma_a+\frac{1}{{\rm Ta}^2}>0,
\ee
which differs from the condition \rf{bsc1} only by the factor $1-\gamma_a\hat\Theta$ at the first term. The reason of this discrepancy is that in \cite{EM1980} $g$ is assumed to be a constant radial gravitational acceleration whereas in \cite{MYM2015} it is the centrifugal acceleration that depends on radius. Hence, substitution of $g=\Omega^2 r$ into Eq. \rf{emsac} is not justified. On the other hand, the factor  $1-\gamma_a\hat\Theta$ is small in the Boussinesq approximation and consequently the equations \rf{bsc1} and \rf{emsacf} can be considered as equivalent.

Taking into account that in the Boussinesq approximation the squared Brunt-V\"ais\"al\"a frequency is ${\rm N}^2=-2\gamma_a\hat\Theta {\rm Rt} \Omega^2$, we transform \rf{emsacf} as
\be{emsacbvf}
4({\rm Ro}+1)+{\rm Pr}\frac{\rm N^2}{\Omega^2}+\frac{1}{{\rm Ta}^2}>0.
\ee
With the opposite sign the inequality  \rf{emsacbvf} is the onset of the double-diffusive Goldreich-Schubert-Fricke instability in the form derived in \citep{AG1978}.

\bibliographystyle{jfm}
\bibliography{KM-JFM-submission}

\begin{thebibliography}{37}
\expandafter\ifx\csname natexlab\endcsname\relax\def\natexlab#1{#1}\fi
\def\au#1{#1} \def\ed#1{#1} \def\yr#1{#1}\def\at#1{#1}\def\jt#1{\textit{#1}}
  \def\bt#1{#1}\def\bvol#1{\textbf{#1}} \def\vol#1{#1} \def\pg#1{#1}
  \def\publ#1{#1}\def\arxiv#1{#1}\def\org#1{#1}\def\st#1{\textit{#1}}

\bibitem[Acheson \& Gibbons(1978)]{AG1978}
{\sc \au{Acheson, D.~J.} \& \au{Gibbons, M.~P.}} \yr{1978}  \at{On the
  instability of toroidal magnetic fields and differential rotation in stars}.
  \jt{Philosophical Transactions of the Royal Society of London A:
  Mathematical, Physical and Engineering Sciences}  \bvol{289}~(1363),
  \pg{459--500}.

\bibitem[Ali \& Weidman(1990)]{Ali}
{\sc \au{Ali, M.} \& \au{Weidman, P.~D.}} \yr{1990}  \at{{On the stability of
  circular Couette flow with radial heating}}.  \jt{Journal of Fluid Mechanics}
   \bvol{220},  \pg{53--84}.

\bibitem[Allilueva \& Shafarevich(2015)]{AS2015}
{\sc \au{Allilueva, A.~I.} \& \au{Shafarevich, A.~I.}} \yr{2015}
  \at{{Asymptotic solutions of linearized Navier--Stokes equations localized in
  small neighborhoods of curves and surfaces}}.  \jt{Russian Journal of
  Mathematical Physics}  \bvol{22}~(4),  \pg{421--436}.

\bibitem[Auer {\em et~al.\/}(1995)Auer, Busse \& Clever]{Auer1995}
{\sc \au{Auer, M.}, \au{Busse, F.~H.} \& \au{Clever, R.~M.}} \yr{1995}
  \at{Three-dimensional convection driven by centrifugal buoyancy}.
  \jt{Journal of Fluid Mechanics}  \bvol{301},  \pg{371--382}.

\bibitem[Balbus \& Potter(2016)]{BP2016}
{\sc \au{Balbus, S.~A.} \& \au{Potter, W.~J.}} \yr{2016}  \at{Surprises in
  astrophysical gasdynamics}.  \jt{Rep. Prog. Phys}  \bvol{79},  \pg{066901}.

\bibitem[Bilharz(1944)]{B1944}
{\sc \au{Bilharz, H.}} \yr{1944}  \at{{Bemerkung zu einem Satze von Hurwitz}}.
  \jt{ZAMM - Journal of Applied Mathematics and Mechanics / Zeitschrift f\"ur
  Angewandte Mathematik und Mechanik}  \bvol{24}~(2),  \pg{77--82}.

\bibitem[Chandrasekhar(1961)]{Chandrasekhar}
{\sc \au{Chandrasekhar, S.}} \yr{1961} {\em Hydrodynamic and hydromagnetic
  stability\/}.  \publ{Oxford: Clarendon}.

\bibitem[Child {\em et~al.\/}(2015)Child, Kersal\'e \& Hollerbach]{CKH2015}
{\sc \au{Child, A.}, \au{Kersal\'e, E.} \& \au{Hollerbach, R.}} \yr{2015}
  \at{{Nonaxisymmetric linear instability of cylindrical magnetohydrodynamic
  Taylor-Couette flow}}.  \jt{Phys. Rev. E}  \bvol{92},  \pg{033011}.

\bibitem[Dobrokhotov \& Shafarevich(1992)]{DS92}
{\sc \au{Dobrokhotov, S.~Yu.} \& \au{Shafarevich, A.~I.}} \yr{1992}
  \at{{Parametrix and the asymptotics of localized solutions of the
  Navier-Stokes equations in R3, linearized on a smooth flow}}.
  \jt{Mathematical Notes}  \bvol{51}~(1),  \pg{47--54}.

\bibitem[Dubrulle {\em et~al.\/}(2005)Dubrulle, Dauchot, Daviaud, Longaretti,
  Richard \& Zahn]{Dubrulle}
{\sc \au{Dubrulle, B.}, \au{Dauchot, O.}, \au{Daviaud, F.}, \au{Longaretti,
  P.-Y.}, \au{Richard, D.} \& \au{Zahn, J.-P.}} \yr{2005}  \at{{Stability and
  turbulent transport in Taylor-Couette flow from analysis of experimental
  data}}.  \jt{Physics of Fluids}  \bvol{17}~(9).

\bibitem[Eckhardt \& Yao(1995)]{EY1995}
{\sc \au{Eckhardt, B.} \& \au{Yao, D.}} \yr{1995}  \at{{Local stability
  analysis along Lagrangian paths}}.  \jt{Chaos, Solitons \& Fractals}
  \bvol{5}~(11),  \pg{2073 -- 2088}.

\bibitem[Eckhoff(1981)]{E1981}
{\sc \au{Eckhoff, K.~S.}} \yr{1981}  \at{On stability for symmetric hyperbolic
  systems, i}.  \jt{Journal of Differential Equations}  \bvol{40}~(1),  \pg{94
  -- 115}.

\bibitem[Economides \& Moir(1980)]{EM1980}
{\sc \au{Economides, D.~G.} \& \au{Moir, G.}} \yr{1980}  \at{{Taylor vortices
  and the Goldreich-Schubert instability}}.  \jt{Geophysical \& Astrophysical
  Fluid Dynamics}  \bvol{16}~(1),  \pg{299--317}.

\bibitem[Friedlander \& Vishik(1995)]{FV1995}
{\sc \au{Friedlander, S.} \& \au{Vishik, M.~M.}} \yr{1995}  \at{On stability
  and instability criteria for magnetohydrodynamics}.  \jt{Chaos}
  \bvol{5}~(2),  \pg{416--423}.

\bibitem[Kirillov \& Verhulst(2010)]{KV2010}
{\sc \au{Kirillov, O.N.} \& \au{Verhulst, F.}} \yr{2010}  \at{{Paradoxes of
  dissipation-induced destabilization or who opened Whitney's umbrella?}}
  \jt{ZAMM - Journal of Applied Mathematics and Mechanics / Zeitschrift f\"ur
  Angewandte Mathematik und Mechanik}  \bvol{90}~(6),  \pg{462--488}.

\bibitem[Kirillov(2013)]{K2013}
{\sc \au{Kirillov, O.~N.}} \yr{2013} {\em {Nonconservative Stability Problems
  of Modern Physics}\/}.  \publ{Berlin, Boston: De Gruyter}.

\bibitem[Kirillov(2016)]{K2016}
{\sc \au{Kirillov, O.~N.}} \yr{2016}  \at{{Singular diffusionless limits of
  double-diffusive instabilities in magnetohy- drodynamics}}.  \jt{Proceedings
  of the Royal Society of London A}  \bvol{subm},  \pg{arXiv:1610.06970v1}.

\bibitem[Kirillov \& Stefani(2013)]{KS2013}
{\sc \au{Kirillov, O.~N.} \& \au{Stefani, F.}} \yr{2013}  \at{Extending the
  range of the inductionless magnetorotational instability}.  \jt{Phys. Rev.
  Lett.}  \bvol{111},  \pg{061103}.

\bibitem[Kirillov {\em et~al.\/}(2014)Kirillov, Stefani \& Fukumoto]{KSF2014}
{\sc \au{Kirillov, O.~N.}, \au{Stefani, F.} \& \au{Fukumoto, Y.}} \yr{2014}
  \at{Local instabilities in magnetized rotational flows: a short-wavelength
  approach}.  \jt{Journal of Fluid Mechanics}  \bvol{760},  \pg{591--633}.

\bibitem[Kucherenko \& Kryvko(2013)]{KK2013}
{\sc \au{Kucherenko, V.~V.} \& \au{Kryvko, A.}} \yr{2013}  \at{{Interaction of
  Alfv\'en waves in the linearized system of magnetohydrodynamics for an
  incompressible ideal fluid}}.  \jt{Russian Journal of Mathematical Physics}
  \bvol{20}~(1),  \pg{56--67}.

\bibitem[Lappa(2012)]{Lappa}
{\sc \au{Lappa, M.}} \yr{2012} {\em Rotating Thermal Flows in Natural and
  Industrial Processes\/}.  \publ{London: John Wiley \& Sons}.

\bibitem[Lifschitz(1991)]{L1991}
{\sc \au{Lifschitz, A.}} \yr{1991}  \at{Short wavelength instabilities of
  incompressible three-dimensional flows and generation of vorticity}.
  \jt{Physics Letters A}  \bvol{157}~(8),  \pg{481 -- 487}.

\bibitem[Lifschitz \& Hameiri(1991)]{LH1991}
{\sc \au{Lifschitz, A.} \& \au{Hameiri, E.}} \yr{1991}  \at{Local stability
  conditions in fluid dynamics}.  \jt{Physics of Fluids A}  \bvol{3}~(11),
  \pg{2644--2651}.

\bibitem[Lifschitz \& Hameiri(1993)]{LH1993}
{\sc \au{Lifschitz, A.} \& \au{Hameiri, E.}} \yr{1993}  \at{Localized
  instabilities of vortex rings with swirl}.  \jt{Communications on Pure and
  Applied Mathematics}  \bvol{46}~(10),  \pg{1379--1408}.

\bibitem[Lifschitz {\em et~al.\/}(1996)Lifschitz, Suters \& Beale]{LSB1996}
{\sc \au{Lifschitz, A.}, \au{Suters, W.~H.} \& \au{Beale, J.~T.}} \yr{1996}
  \at{{The Onset of Instability in Exact Vortex Rings with Swirl}}.
  \jt{Journal of Computational Physics}  \bvol{129}~(1),  \pg{8 -- 29}.

\bibitem[Lifshitz(1987)]{L1987}
{\sc \au{Lifshitz, A.~E.}} \yr{1987}  \at{{Continuous spectrum in general
  toroidal systems (ballooning and Alfv\'en modes})}.  \jt{Physics Letters A}
  \bvol{122}~(6),  \pg{350 -- 356}.

\bibitem[Lopez {\em et~al.\/}(2013)Lopez, Marques \& Avila]{Lopez2013}
{\sc \au{Lopez, J.~M.}, \au{Marques, F.} \& \au{Avila, M.}} \yr{2013}  \at{{The
  Boussinesq approximation in rapidly rotating flows}}.  \jt{J. Fluid Mech.}
  \bvol{737},  \pg{56--77}.

\bibitem[Marcus {\em et~al.\/}(2013)Marcus, Pei, Jiang \&
  Hassanzadeh]{Zombie2013}
{\sc \au{Marcus, P.~S.}, \au{Pei, S.}, \au{Jiang, C.-H.} \& \au{Hassanzadeh,
  P.}} \yr{2013}  \at{Three-dimensional vortices generated by self-replication
  in stably stratified rotating shear flows}.  \jt{Phys. Rev. Lett.}
  \bvol{111},  \pg{084501}.

\bibitem[Maslov(1986)]{Maslov1986}
{\sc \au{Maslov, V.~P.}} \yr{1986}  \at{{Coherent structures, resonances, and
  asymptotic non-uniqueness for Navier-Stokes equations with large Reynolds
  numbers}}.  \jt{Russian Mathematical Surveys}  \bvol{41}~(6),  \pg{23--42}.

\bibitem[Meyer {\em et~al.\/}(2015)Meyer, Yoshikawa \& Mutabazi]{MYM2015}
{\sc \au{Meyer, A.}, \au{Yoshikawa, H.~N.} \& \au{Mutabazi, I.}} \yr{2015}
  \at{{Effect of the radial buoyancy on a circular Couette flow}}.  \jt{Physics
  of Fluids}  \bvol{27}~(11),  \pg{114104}.

\bibitem[Mutabazi \& Bahloul(2002)]{MUBA2002}
{\sc \au{Mutabazi, I.} \& \au{Bahloul, A.}} \yr{2002}  \at{Stability analysis
  of a vertical curved channel flow with a radial temperature gradient}.
  \jt{Theoretical and Computational Fluid Dynamics}  \bvol{16}~(1),
  \pg{79--90}.

\bibitem[Nelson {\em et~al.\/}(2013)Nelson, Gressel \& Umurhan]{NGU2013}
{\sc \au{Nelson, R.~P.}, \au{Gressel, O.} \& \au{Umurhan, O.~M.}} \yr{2013}
  \at{Linear and non-linear evolution of the vertical shear instability in
  accretion discs}  \bvol{435}~(3),  \pg{2610--2632}.

\bibitem[Stefani \& Kirillov(2015)]{SK2015}
{\sc \au{Stefani, F.} \& \au{Kirillov, O.~N.}} \yr{2015}  \at{Destabilization
  of rotating flows with positive shear by azimuthal magnetic fields}.
  \jt{Phys. Rev. E}  \bvol{92},  \pg{051001}.

\bibitem[Tuckerman(2001)]{LT2001}
{\sc \au{Tuckerman, L.~S.}} \yr{2001}  \at{{Thermosolutal and binary fluid
  convection as a $2 \times 2$ matrix problem}}.  \jt{Physica D}  \bvol{156},
  \pg{325--363}.

\bibitem[Urpin \& Brandenburg(1998)]{UB1998}
{\sc \au{Urpin, V.} \& \au{Brandenburg, A.}} \yr{1998}  \at{Magnetic and
  vertical shear instabilities in accretion discs}.  \jt{Monthly Notices of the
  Royal Astronomical Society}  \bvol{294}~(3),  \pg{399--406}.

\bibitem[Vinberg(2003)]{V2003}
{\sc \au{Vinberg, E.~B.}} \yr{2003} {\em {A course in algebra}\/}.
  \publ{American Mathematical Society: Providence, R.I}.

\bibitem[Yoshikawa {\em et~al.\/}(2013)Yoshikawa, Nagata \& Mutabazi]{YNM2013}
{\sc \au{Yoshikawa, H.~N.}, \au{Nagata, M.} \& \au{Mutabazi, I.}} \yr{2013}
  \at{Instability of the vertical annular flow with a radial heating and
  rotating inner cylinder}.  \jt{Physics of Fluids}  \bvol{25}~(11),
  \pg{114104}.

\end{thebibliography}

\end{document}